\documentclass[conference]{IEEEtran}
\usepackage{times}
\usepackage{xcolor}
\usepackage{xfrac}
\usepackage{soul}
\usepackage[utf8]{inputenc}
\usepackage{amsmath}
\usepackage[
  separate-uncertainty = true,
  multi-part-units = repeat
]{siunitx}
\usepackage{tabstackengine}
\stackMath
\usepackage{amsfonts,amssymb}
\usepackage[small]{caption}
\usepackage{graphicx}
\usepackage{amsfonts}
\usepackage{multirow}
\usepackage{subcaption}
\graphicspath{{figs/}} 
\newcommand{\sys}{DeepMarks}

\begin{document}
\title{DeepMarks: A Digital Fingerprinting Framework for Deep Neural Networks}
\author{
{Huili Chen, Bita Darvish Rohani,
and Farinaz Koushanfar}
\\ 
University of California, San Diego  \\
huc044@ucsd.edu, bita@ucsd.edu
farinaz@ucsd.edu
}

\maketitle
\begin{abstract}
This paper proposes \sys{}, a novel end-to-end framework for systematic fingerprinting in the context of Deep Learning (DL). Remarkable progress has been made in the area of deep learning. Sharing the trained DL models has become a trend that is ubiquitous in various fields ranging from biomedical diagnosis to stock prediction. As the availability and popularity of pretrained models are increasing, it is critical to protect the Intellectual Property (IP) of the model owner. \sys{} introduces the first fingerprinting methodology that enables the model owner to embed unique fingerprints within the parameters (weights) of her model and later identify undesired usages of her distributed models. The proposed framework embeds the fingerprints in the Probability Density Function (pdf) of trainable weights by leveraging the extra capacity available in contemporary DL models. \sys{} is robust against fingerprints collusion as well as network transformation attacks, including model compression and model fine-tuning. Extensive proof-of-concept evaluations on MNIST and CIFAR10 datasets, as well as a wide variety of deep neural networks architectures such as Wide Residual Networks (WRNs) and Convolutional Neural Networks (CNNs), corroborate the effectiveness and robustness of \sys{} framework. 
\end{abstract}

\vspace{0.5em}
\IEEEpeerreviewmaketitle
\section{Introduction}   \label{intro}
The recent advance in deep learning and neural networks has provided a paradigm shift in various scientific fields. In particular, numerous deep neuron networks (DNNs) such as GoogLeNet~\cite{GooLeNet}, AlexNet~\cite{AlexNet}, Residual Network~\cite{ResNet}, and Neural
Architecture Search networks~\cite{zoph2017learning} have become prevalent standards for applications including autonomous transportation, automated manufacturing, natural language processing, intelligent warfare and smart health~\cite{lecun2015deep, schmidhuber2015deep, collobert2008unified}. Meanwhile, open-sourced deep learning frameworks have enabled users to develop customized machine learning systems based on the existing models. PyTorch~\cite{PyTorch}, Tensorflow~\cite{abadi2016tensorflow}, Keras~\cite{chollet2015keras}, MXNet~\cite{chen2015mxnet}, and Caffe~\cite{jia2014caffe} are examples of such tools.

The distribution of pre-trained neural networks is a promising trend and makes the utilization of DNNs easier. For instance, Caffe provides Model Zoo that includes built neural networks and pre-trained weights for various applications~\cite{caffe_modelZoo}. As the accessibility of models increases, a practical concern is the IP protection and Digital Right Management (DRM) of the distributed models. On the one hand, DL models are usually trained by allocating significant computational resources to process massive training data. The built models are therefore considered as the owner's IP and need to be protected to preserve the competitive advantages. On the other hand, malicious attackers may take advantage of the models for illegal usages. The potential problems need to be taken into account during the design and training of the DL models before the owners make their models publicly available.

Previous works have identified the importance of IP protection in DL domain and propose watermarking methodologies for DNNs. The authors of~\cite{uchida2017embedding, nagai2018digital} present a new approach for watermarking DNNs by embedding the IP information in the weights. The embedded watermark can be extracted by the owner assuming the details of the models are available to the owner (`white-box' setting). To provide IP protection for a remote neural network where the model is exposed as a service (`black-box' setting), the paper~\cite{merrer2017adversarial} proposes a zero-bit watermarking methodology by tweaking the decision boundary. The paper~\cite{DeepSigns} presents a generic watermarking framework for IP protection in both white-box and black-box scenarios by embedding the watermarks in the pdf of the activation sets of the target layers. To the best of our knowledge, there is no prior work that has targeted fingerprinting for DNNs.  

This paper proposes \sys{}, a novel end-to-end framework that enables coherent integration of robust digital fingerprinting in contemporary deep learning models. \sys{}, for the first time, introduces a \textit{generic} functional fingerprinting methodology for DNNs. The proposed methodology is simultaneously \textit{user and model dependent}. \sys{} works by assigning a unique binary code-vector (a.k.a., \textit{fingerprint}) to each user and embedding the fingerprint information in the probabilistic distribution of the weights while preserving the accuracy. We demonstrate the robustness of our proposed framework against collusion and transformation attacks including model compression/pruning, and model fine-tuning. The explicit technical contributions of this paper are as follows:

\begin{itemize}
\item Proposing \sys{}, the first end-to-end framework for systematic deep learning IP protection and digital right management. A novel fingerprinting methodology is introduced to encode the pdf of the DL models and effectively trace the IP ownership as well as the usage of each distributed model. 
        
\item Introducing a comprehensive set of qualitative and quantitative metrics to assess the performance of a fingerprinting methodology for (deep) neural networks. Such metrics provide new perspectives for model designers and enable coherent comparison of current and pending DL IP protection techniques. 
    
\item Performing extensive proof-of-concept evaluation on various benchmarks including commonly used MNIST, CIFAR10 datasets. Our evaluations corroborate the effectiveness of \sys{} to detect IP ownership and track the individual culprits/colluders who use the model for unintended purposes. 
\end{itemize}

\begin{table*}[ht!]
\centering
\caption{Requirements for an effective fingerprinting methodology of deep neural networks.}
\label{tab:required}
\scalebox{0.98}{
\begin{tabular}{|l||p{14cm}|}
\hline
\multicolumn{1}{|c||}{\textbf{Requirements}}   & \multicolumn{1}{|c|}{\textbf{Description}}                                                \\ \hline \hline
Fidelity     & The functionality (e.g., accuracy) of the host neural network shall not be degraded as a result of fingerprints embedding.    
\\ \hline
Uniqueness  & The fingerprint need to be unique for each user, which enables the owner to trace the unintended usage of the distributed model conducted by any specific user. 
\\ \hline
Capacity     & The fingerprinting methodology shall be capable of embedding a large amount of information in the host neural network.                           \\ \hline 
Efficiency    & The overhead of fingerprints embedding and extraction shall be negligible.   \\ \hline 
Security     & The fingerprint shall leave no tangible footprint in the host neural network; thus, an unauthorized individual cannot detect the presence of a fingerprint in the model. 
\\ \hline
Robustness   & The fingerprinting methodology shall be resilient against model modifications such as compression/pruning, fine-tuning. Furthermore, the fingerprints shall be resistant to collusion attacks where the adversaries try to produce an unmarked neural network using multiple marked models. 
\\ \hline
Reliability  & The fingerprinting methodology should yield minimal false negatives, suggesting that the embedded fingerprint should be detected with high probability. 
\\ \hline
Integrity    & The fingerprinting methodology should yield minimal false alarm (a.k.a., false positive). This means that the probability of an innocent user being accused as a colluder should be very low. 
\\ \hline
Scalability & The fingerprinting methodology should be able to support a large number of users because of the nature of the model distribution and sharing.\\ \hline

Generality& The fingerprinting methodology should be applicable to various neural network architectures and datasets.  \\ \hline
\end{tabular}}
\end{table*}

\section{Problem Formulation}  \label{prob_form}
Fingerprinting is defined as the task of embedding a $v$-bit binary code-vector $\mathbf{c_j} \in \{0,1\}^v$ in the weights of a host neural network. Here, $j=1,...,n$ denotes the index for each distributed user where $n$ is the number of total users. The fingerprint information can be either embedded in one or multiple layers of the DNN model. The objective of fingerprinting is two-fold: (i) claiming the ownership of a specific neural network, and (ii) tracing the unintended usage of the model by distributed users. In the following sections, we formulate the requirements for digital fingerprinting in the context of DL and discuss possible attacks that might render the embedded fingerprints ineffective.

\subsection{Requirements}  \label{reqiurements}
Table~\ref{tab:required} summarizes the requirement for an effective fingerprints in the deep neural network domain. In addition to fidelity, efficiency, security, capacity, reliability, integrity, and robustness requirements that are shared between fingerprinting and watermarking, a successful fingerprinting methodology should also satisfy the uniqueness, scalability, and collusion resilience criteria.   

On the one hand, \textbf{uniqueness} is the intrinsic property of fingerprints. Since the model owner aims to track the usage of the model distributed to each specific user, the uniqueness of fingerprints is essential to ensure correct identification of the target user.  On the other hand, as the number of participants involved in the distribution of neural networks increases, \textbf{scalability} is another key factor to perform IP protection and digital right management in large-scale settings. Particularly, the fingerprinting methodology should be able to accommodate a large number of distributed users.   

Collusion attacks can result in the attenuation of the fingerprint from each colluder and have been identified as cost-effective attacks in the multi-media domain. In a traditional collusion attack, multiple users work together to produce an unmarked content using differently marked versions of the same content~\cite{wu2004collusion}. In the domain of DL, a group of users who have the same host neural network but different fingerprints may work collaboratively to construct a model where no fingerprints can be detected by the owner. Considering the practicality of such attacks, We include \textbf{collusion resilience} in the robustness requirement for DNN fingerprinting.

\subsection{Attack Models}  \label{attacks}
Corresponding to the robustness requirements listed in Table~\ref{tab:required}, we discuss three types of DL domain-specific attacks that the fingerprinting methodology should be resistant to: model fine-tuning, model compression, and collusion attacks.

\vspace{0.3em}
\noindent \textbf{Model Fine-tuning.} Fine-tuning the pre-trained neural networks for transfer learning is a common practice since training a DL model from scratch is computationally expensive~\cite{shin2016deep}. For this reason, model fine-tuning can be an unintentional model transformation conducted by honest users or an intentional attack performed by malicious users. The parameters of the model are changed during fine-tuning, therefore the embedded fingerprints should be robust against this modification. 

\vspace{0.3em}
\noindent \textbf{Model Compression.} Compressing the DNN models by parameter pruning is a typical technique to reduce the computational overhead of executing a neural network~\cite{han2015learning}. Genuine users may leverage parameter pruning to make their models compressed while adversaries may apply pruning to remove the fingerprints embedded by the owner. Since pruning alters the model parameters that carry the fingerprints information, an effective fingerprinting methodology shall be resistant to parameter pruning. 

\vspace{0.3em}
\noindent \textbf{Collusion Attack.} Multiple attackers who have the same host neural network with different embedded fingerprints may perform collusion attacks to produce an unmarked model. We consider fingerprints averaging attack which is a common collusion attack and demonstrate how \sys{} is robust against such attacks. 

\begin{figure*}[ht!]
\centering
 \includegraphics[width=0.9\textwidth]{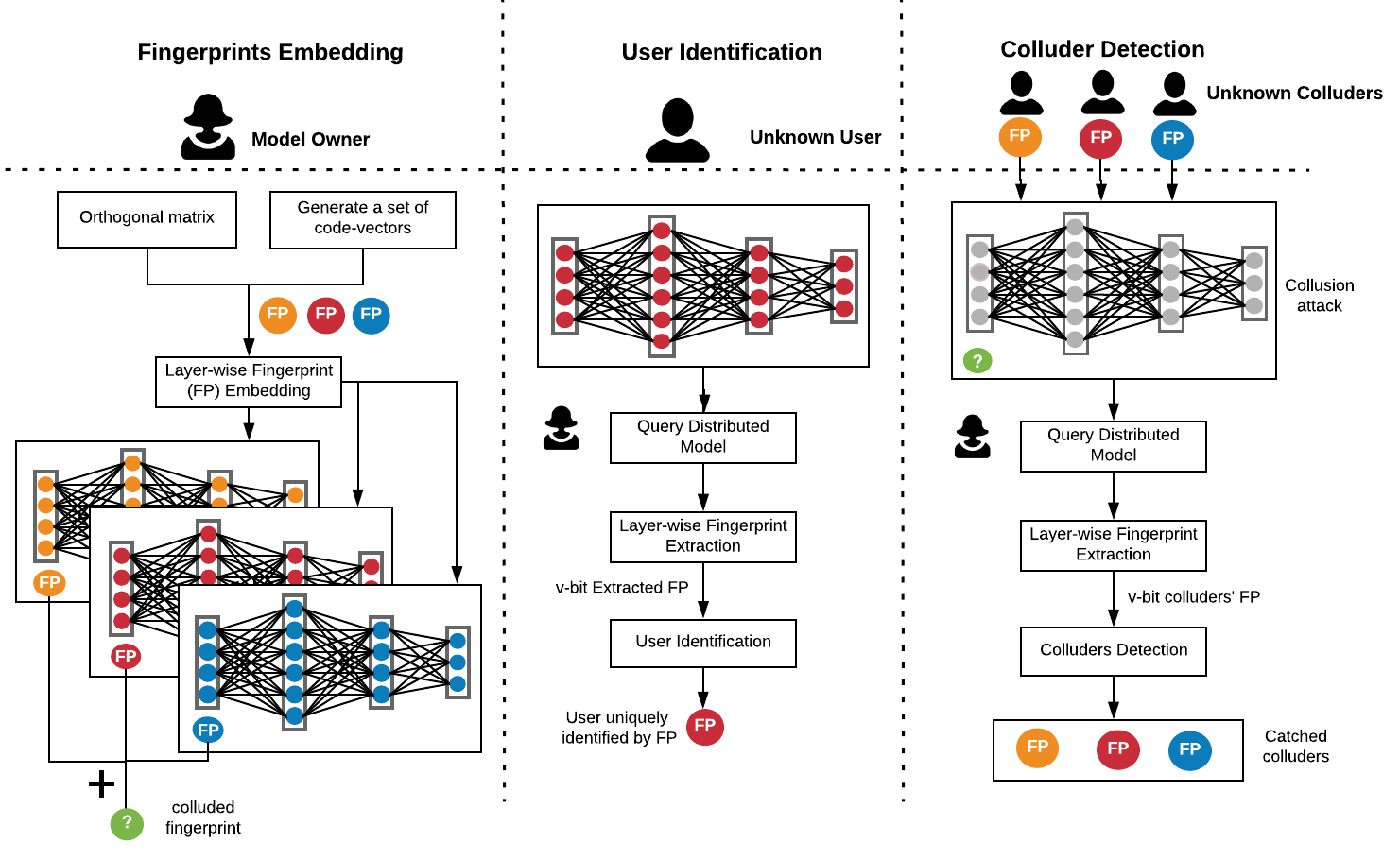} 
\caption{\label{fig:global} \sys{} Global Flow: \sys{} performs fingerprinting on DL models by embedding the designated fingerprinting information in the distribution of weights for selected layers. To enable IP protection and digital right management, \sys{} allows the model owner to extract the embedded fingerprints for user identification as well as colluder detection after distributing the models.}
\end{figure*}

\section{Fingerprint Embedding}  \label{fp_embed}
The global flow of \sys{} is illustrated in Figure~\ref{fig:global}. In order to trace the models that are distributed to individual users, the owner first assigns a specific code-vector to each user. Given the code-vector and an orthonormal basis matrix, a unique fingerprint is constructed to identify each user. The designed fingerprint is then embedded in the weights distribution for each user by fine-tuning the model with an additive \textit{embedding loss}. To identify a specific user, the owner assesses the weights of the marked layers in her model and extracts the corresponding code-vector. The decoded code-vector thus uniquely identifies the inquired user. In addition, \sys{} enables the owner to detect colluders who work collaboratively and try to generate a model where no fingerprints can be detected by the owner. 

There are two types of fingerprint modulation mechanisms in the multi-media domain: (i) \textbf{orthogonal modulation}, and (ii) \textbf{coded modulation}~\cite{wu2004collusion}. In the rest of this section, we discuss how \sys{} framework adopts these two fingerprinting methods to provide a generic solution for DNNs.

\subsection{Orthogonal Fingerprinting}  \label{orthog_embed}
As discussed in Section~\ref{reqiurements}, uniqueness is an essential requirement for fingerprinting to track individual users. \textit{Orthogonal modulation} is a technique that uses orthogonal signals to represent different information~\cite{wu2003data}. By using mutually orthogonal watermarks as fingerprints, the separability between users can be maximized. Given an orthogonal matrix $\mathbf{U}_{v \times v} = [\mathbf{u_1}, ..., \mathbf{ u_v}]$, the unique fingerprint for user $j$ can be constructed by assigning each column to a user:
\begin{equation}  \label{eq:orthog_fp}
	 \mathbf{ f_j}= \mathbf{u_j},
\end{equation}
where $\mathbf{u_j}$ is the $j^{th}$ column of the matrix $\mathbf{U}$, $j=1,...v$. Here, $v$ orthogonal signals deliver $B=log_2 v$ bits information and can be recovered from $v$ correlators. The orthogonal matrix can be generated from element-wise Gaussian distribution~\cite{wu2004collusion}.

Regularizing neural networks for security purpose has been presented in previous works~\cite{CuRTAIL, DeepSigns}. However, none of these works focuses on the fingerprinting of the DL models. \sys{} embeds the constructed fingerprint in the target layers of the host model by adding the following term to the loss function conventionally used for training/fine-tuning deep neural networks: 
\begin{equation}  \label{eq:embed_loss}
\mathcal{L} = \mathcal{L}_0 \; + \gamma~MSE(\mathbf{f_j} - \mathbf{X}\mathbf{w}).
\end{equation}
Here, $\mathcal{L}_0$ is the conventional loss function (e.g. cross-entropy loss), $MSE$ is the mean square error function, $\gamma$ is the embedding strength that controls the trade-off between the two loss terms, $\mathbf{X}$ is the secret random projection matrix generated by the owner. $\mathbf{w}$ is the flattened averaged weights of the target layers for embedding the pertinent fingerprint. 

As a proof-of-concept analysis, we embed the fingerprint $\mathbf{f_j}$ in the convolutional layer of the host neural network, thus the weight $\mathbf{W}$ is a 4D tensor $\mathbf{W} \in \mathbb{R}^{D \times D \times F \times H}$ where $D$ is the input depth, $F$ is the kernel size, and $H$ is the number of channels in the convolutional layer. The ordering of filter channels does not change the output of the neural network if the parameters in the consecutive layers are rearranged correspondingly~\cite{uchida2017embedding}. As such, we take the average of $\mathbf{W}$ over all channels and stretch the resulting tensor into a vector $\mathbf{w} \in \mathbb{R}^N$, where $N = D \times D \times F$. The rearranged weight vector $\mathbf{w}$ is then multiplied with a secret random matrix $\mathbf{X} \in \mathbb{R}^{v \times N}$ and compared with the fingerprint $\mathbf{f_j}$. The additional \textit{embedding loss} term $MSE(\mathbf{f_j} - \mathbf{Xw})$ inserts the fingerprint $\mathbf{f_j}$ in the distribution of the target layer weights by enforcing the model to minimize the embedding loss together with the conventional loss during the training/fine-tuning of the DNN model. 

Since each user corresponds to a column vector in the orthogonal basis matrix, the maximum number of users is equal to the dimension of the fingerprint (which is also the number of orthogonal bases): $n=v$. Thus, the amount of customers that the same neural network can be distributed to is limited by the fingerprints dimension. Orthogonal fingerprints are developed based on spread spectrum watermarking~\cite{cox1997secure}. The straightforward concept and simplicity of implementation make orthogonal fingerprinting attractive to identification applications where only a small group of users are involved. Although orthogonality helps to distinguish individual users, the independent nature of orthogonal fingerprints makes it vulnerable to collusion attacks~\cite{wu2004collusion}.

\subsection{Coded Fingerprinting} \label{code_embed}
To support a large group of users and improve the collision resilience of the fingerprints, coded modulation is leveraged to introduce correlation between fingerprints~\cite{trappe2003anti,trappe2002collusion}. Similar ideas have been discussed in Antipodal CDMA-type watermarking where the correlation contributions only decrease at the locations where the watermarks code-bits are different~\cite{liu2005multimedia}. Correlation not only allows the system to support a larger number of fingerprints than the dimensionality of the orthogonal basis vectors, but also alleviates the attenuation of fingerprints due to collusion attacks. The challenge for coded fingerprinting is to design code-vectors such that (i) the correlations are introduced in a strategical way, and (ii) the correct identification of the users involved in a collusion attack is facilitated.

Anti-collusion codes (ACC) is proposed in~\cite{wu2004collusion} for coded fingerprinting and have the property that the composition of any subset of $K$ or fewer code-vectors is unique. This property allows the owner to identify a group of $K$ or fewer colluders accurately. A $K$-resilient \textit{AND-ACC} is a codebook where the element-wise composition is logic-AND and allows for the accurate identification of $K$ unique colluders from their composition. Previous works in the multi-media domain have shown that \textit{Balanced Incomplete Block Design} (BIBD) can be used to generate ACCs of binary values~\cite{yu2010group}. A $(v,k, \lambda)$-BIBD is a pair $(\mathcal{X}, \mathcal{A})$ where $\mathcal{A}$ is the collection of $k$-element subsets (blocks) of a $v$-dimension set $\mathcal{X}$ such that each pair of elements of $\mathcal{X}$ appear together exactly $\lambda$ times in the subsets~\cite{trappe2003anti, dinitz1992contemporary}. The $(v,k,\lambda)$-BIBD has $b=\sfrac{\lambda(v^2-v)}{(k^2-k)}$ blocks ($k$ is the block size) and can be represented by its corresponding incidence matrix $\mathbf{C}_{v \times b}$. The elements in the incidence matrix have binary values where:
\begin{equation*}
c_{ij} = 
	\begin{cases}
    1, \; \text{if $i^{th}$ value occurs in $j^{th}$ block}  \\ 
    0, \; \text{otherwise}.
    \end{cases}
\end{equation*}

By setting the number of concurrent occurrence to one ($\lambda=1$) and assigning the bit complement of columns of the incidence matrix $\mathbf{C}_{v \times b}$ as the code-vectors, the resulting $(v,k,1)$-BIBD code is $(k-1)$-resilient and supports up to $n=b$ users~\cite{trappe2003anti}. The theory of BIBD shows that the parameters satisfy the relationship $b>v$~\cite{dinitz1992contemporary}, which means the number of users (or fingerprints) is larger than the dimension of the orthogonal basis vectors. More specifically, the BIBD-ACC construction only requires $\mathcal{O}(\sqrt{n})$ basis vectors to accommodate $n$ users instead of $\mathcal{O}(n)$ in orthogonal fingerprinting scheme. Systematic approaches for constructing infinite families of BIBDs have been developed~\cite{colbourn2006handbook}, which provides a vast supply of ACCs. 

Given the designed incidence matrix $\mathbf{C}_{v \times b}$, the coefficient matrix $\mathbf{B}_{v \times b}$ for fingerprints embedding can be computed from the linear mapping $b_{ij} = 2 c_{ij} - 1 $, thus $b_{ij} \in \left\{\pm 1\right \}$ corresponds to the antipolar form~\cite{proakis1994communication}. The fingerprint for $j^{th}$ user is then constructed from the orthogonal matrix $\mathbf{U}_{v \times v}$ and the coefficient matrix $\mathbf{B}_{v \times b}$ as follows:
\begin{equation}  \label{eq:coded_fp}
	 \mathbf{ f_j}= \sum_{i=1}^v b_{ij} \mathbf{u_j},
\end{equation}
where $\mathbf{b_j} \in \left\{ \pm 1 \right \}^v$ is the coefficient vector associated with user $j$. Finally, The designed fingerprint $\mathbf{f_j}$ is embedded in the weights of the target model by adding the embedding loss to the conventional loss as shown in Equation~\ref{eq:embed_loss}. 

Comparing the orthogonal fingerprinting in Equation~\ref{eq:orthog_fp} with the coded fingerprinting in Equation~\ref{eq:coded_fp}, one can see that orthogonal fingerprinting can be implemented by coded fingerprinting if an identity matrix is used as the ACC codebook $\mathbf{C}= \mathbf{I}$. This, in turn, means that the code-vector assigned to each user only has one element that equals to $1$ and all the others are zeros. Therefore, orthogonal fingerprinting can be considered as a special case of coded fingerprinting. 

\section{Fingerprint Extraction} \label{fp_extraction}
For the purpose of fingerprints inquiry and colluder detection, the model owner assesses the weights of the marked layers, recovers the code-vector assigned to the user, and uses correlation statistics (orthogonal fingerprinting method) or BIBD ACC codebook (coded fingerprinting method) to identify colluders. Note that in the multi-media domain, there are two types of detection schemes for spread spectrum fingerprinting: blind or non-blind detection, depending on whether the original host signal is available in the detection stage or not. Non-blind detection has higher confidence in detection while blind detection is applicable in distributed detection settings~\cite{wu2004collusion, zhao2006fingerprint}. \sys{} leverages blind detection scheme and does not require the knowledge of the original content; thus content registration and storage resources are not needed. We discuss the workflow of extracting the code-vector from the marked weights and detecting participants in fingerprints collusion attacks for both fingerprinting methods in the following sections.

\subsection{Orthogonal Fingerprinting}\label{orthog_detect}
\vspace{0.3em}
\noindent \textbf{Code-vector extraction.} As discussed in Section~\ref{prob_form}, one objective of embedding fingerprints in the DNNs is to uniquely identify individual users. Since the fingerprint is determined by the corresponding code-vector, we formulate the problem of user identification as code-vector extraction from the marked weighs in each distributed model. 

The embedding methodology of orthogonal fingerprinting is described in Section~\ref{orthog_embed}. In the inquiry stage, \sys{} first acquires the weights tensor $\mathbf{\widetilde{W_j}}$ of the pertinent marked layers for the target user $j$ and computes the flattened averaged version $\mathbf{\widetilde{w_j}}$. The fingerprint is recovered from the multiplication $\mathbf{\widetilde{f_j}} = \mathbf{X \widetilde{w_j}}$ where $\mathbf{X}$ is the random projection matrix specified by the owner. For simplicity, we use orthonormal columns to construct the basis matrix $\mathbf{U}$, thus the correlation score vector (which is also the coefficient vector) can be computed as follows:
\begin{equation}  \label{eq:orthog_decode}
\mathbf{\widetilde{b_j}} = \mathbf{\widetilde{f_j}}^T \mathbf{U} = [\mathbf{\widetilde{f_j}}^T \mathbf{u_1}, ..., \mathbf{\widetilde{f_j}}^T \mathbf{u_v}].
\end{equation}
Since the fingerprints are orthogonal, only $j^{th}$ component in the correlation scores $\mathbf{\widetilde{b_j}}$ will have large magnitude while all the other elements will be nearly zeros. Finally, the code-vector $\mathbf{\widetilde{c_j}} \in \left\{0, 1\right \}^v$ assigned to $j^{th}$ user is extracted by element-wise hard-thresholding of the correlation vector $\mathbf{\widetilde{b_j}}$.

\vspace{0.5em}
\noindent \textbf{Colluder detection.} 
Recall that the second objective of the owner for leveraging fingerprinting is to trace illegal redistribution or unintended usages of the models. Here we consider a typical linear collusion attack where $K$ colluders average their fingerprints and collaboratively generate a new model where the fingerprint is not detectable. To detect participants in the collusion attack, the owner first computes the correlation scores between the colluded fingerprint and each basis vector as shown in Equation~\ref{eq:orthog_decode}. Element-wise hard-thresholding is then performed on the correlation vector where the positions of ``1"s correspond to the indices of the colluders. According to Equation~\ref{eq:orthog_decode}, the magnitude of averaged fingerprints is attenuated by $\frac{1}{K}$ assuming there are $K$ colluders participating in the attack. As shown in~\cite{wu2004collusion}, $\mathcal{O}(\sqrt{\sfrac{v}{logv}})$ colluders are sufficient to defeat the fingerprinting system, where $v$ is the dimension of the fingerprint.

\subsection{Coded Fingerprinting}  \label{coded_detect}
\vspace{0.5em}
\noindent \textbf{Code-vector extraction.} Similar to the extraction of orthogonal fingerprints, the owner acquires the weights in the marked layers $\mathbf{ \widetilde{W_j} }$ and computes its averaged flattened version $\mathbf{ \widetilde{w_j}}$, then extracts the colluders' fingerprint $\mathbf{ \widetilde{f_j} = X \widetilde{w_j}}$. The extracted fingerprint is then multiplied with the basis matrix to compute the correlation score vector $\mathbf{\widetilde{b_j}} = \mathbf{\widetilde{f_j}}^T \mathbf{U}$. Finally, the ACC code-vector $\mathbf{ \widetilde{c_{j}} }$  assigned to the $j^{th}$ user is decoded from $\mathbf{ \widetilde{b_{j}}}$ by hard-thresholding.

To illustrate the workflow of code-vector extraction for coded fingerprinting, let us consider a $(7,3,1)$-BIBD codebook given in Equation~\ref{eq:codebook_eg}. The coefficient vector of each fingerprint is constructed by mapping each column of the codebook $\mathbf{C}$ to the antipodal form $\left\{ \pm1 \right \}$. The fingerprints for all users are shown in Equation~\ref{eq:code_fp_eg}:
\begin{equation}  \label{eq:codebook_eg}
    \mathbf{C} = 
    \begin{pmatrix}
    0 &0 &0 &1 &1 &1 &1 \\
    0 &1 &1 &1 &0 &1 &1 \\
    1 &0 &1 &0 &1 &0 &1 \\
    0 &1 &1 &1 &1 &0 &0 \\
    1 &1 &0 &0 &1 &1 &0 \\
    1 &0 &1 &1 &0 &1 &0 \\
    1 &1 &0 &1 &0 &0 &1 
    \end{pmatrix},
\end{equation}
\begin{equation}  \label{eq:code_fp_eg}
  \begin{cases}
    \mathbf{f_1} = -\mathbf{u_1} -\mathbf{u_2} + \mathbf{u_3} -\mathbf{u_4} + \mathbf{u_5} + \mathbf{u_6} + \mathbf{u_7}, \\
    \cdots \\
    \mathbf{f_6} = +\mathbf{u_1} +\mathbf{u_2} - \mathbf{u_3}
  -\mathbf{u_4} + \mathbf{u_5} +\mathbf{u_6}  -\mathbf{u_7},  \\
    \mathbf{f_7} = +\mathbf{u_1} +\mathbf{u_2} +\mathbf{u_3}
  -\mathbf{u_4} -\mathbf{u_5} 
  -\mathbf{u_6}  +\mathbf{u_7},  \\
  \end{cases}
\end{equation}
where $\mathbf{u_i}~(i=1,...,7)$ are orthogonal columns of the matrix $\mathbf{U}$. For user $1$, her coefficient vector can be recovered by computing the correlation scores:
\begin{equation*}
  \mathbf{\widetilde{b_1}} =     \mathbf{f_1}^T[\mathbf{u_1}, ..., \mathbf{u_7}] = [-1, -1, +1, -1, +1, +1, +1].
\end{equation*}
The corresponding code-vector is then extracted by the inverse linear mapping $c_{ij} = \frac{1}{2}(b_{ij}+1)$. The resulting code-vector is $\mathbf{\widetilde{c_1}} = [0, 0, 1, 0, 1, 1, 1]$, which is exactly the same as the first column of $\mathbf{C}$. The consistency shows that BIBD AND-ACC codebooks can be leveraged to identify individual users. 

\vspace{0.5em}
\noindent \textbf{Colluder detection.} Recall that in Section~\ref{code_embed}, we discuss the property of BIBD and its application for constructing anti-collusion codes. Here, we describe how to use the intrinsic asset of AND-ACC for colluder detection in fingerprints averaging attack. Assuming the positions of the marked layer are known to the colluders, they can perform element-wise average on their weight tensors in the pertinent layers and generate $\mathbf{W^{avg}}$ as the response to the owner's inquiry. The owner then computes the correlation vector $\mathbf{b^{avg}}$ as follows:
\begin{align}  \label{eq:coded_avg}
	\mathbf{f^{avg}} &= \mathbf{X w^{avg}},   \\
     \mathbf{ b^{avg} } &= \mathbf{ (f^{avg})^T }\mathbf{U}.
\end{align}

The problem of identifying colluders based on the correlation statistics has been well addressed in conventional fingerprinting that is based on spread spectrum watermarking~\cite{cox1997secure, jain2000digital}. There are three main schemes: hard-thresholding detector, adaptive sorting detector, and sequential detector~\cite{wu2004collusion}. Hard-thresholding detector works by comparing each element in the correlation score vector $\mathbf{b}$ with a threshold $\tau$ to decide the corresponding bit (``0" or ``1") in the ACC code-vector. Adaptive sorting detector sorts the correlation scores in a descending order and iteratively narrow down the set of suspected users until the corresponding likelihood estimation of the colluder set stops increasing. Sequential detector directly estimates the colluder set from the pdf of the correlation statistics without decoding the ACC code-vector. For details about each detection method, we refer readers to the paper~\cite{wu2004collusion}.

\sys{} deploys hard-thresholding detector for colluders identification. The ACC code-vector is decoded from the correlation vector $\mathbf{b^{avg} } = [b^{avg}_1, ..., b^{avg}_v]$ by comparing each component with the threshold $\tau$:
\begin{equation}  \label{eq:b_threshold}
	c^{avg}_i= 
    \begin{cases}
		1, \text{if $b^{avg}_i > \tau$}, \\
        0, \text{otherwise}.
	\end{cases}
\end{equation}
Given the ACC code-vector of the colluders $\mathbf{c^{avg}}$, the remaining problem is to find the subsets of columns from the codebook $\mathbf{C}$ such that their logic-AND composition is equal to $\mathbf{c^{avg}}$. For a $(v,k,1)$-BIBD-ACC, at most $(k-1)$ colluders can be uniquely identified. 

As an example, we demonstrate the colluder detection scheme of \sys{} using the $(7,3,1)$-BIBD codebook given in Equation~\ref{eq:codebook_eg}. Assuming user $6$ and user $7$ collectively generate the averaged fingerprint:
\begin{align*}
  \mathbf{f^{avg}} &= \frac{1}{2}(\mathbf{f_6}+\mathbf{f_7}),  \\ 
  &= \frac{1}{2} (2\mathbf{u_1}+2\mathbf{u_2}-2\mathbf{u_4}).
\end{align*}
The owner assesses the averaged fingerprint and computes the correlation scores as the following:
\begin{align*}
	\mathbf{b^{avg}}&= \mathbf{(f^{avg})^T} \mathbf{U} = [1, 1, 0, -1, 0, 0, 0].
\end{align*}
The colluders' code-vector is then extracted according to decision rule in Equation~\ref{eq:b_threshold}:
\begin{equation*}
	\mathbf{c^{avg}} = [1, 1, 0, 0, 0, 0, 0].
\end{equation*}
It can be observed that the logic-AND of column 6 and column 7 in the codebook $\mathbf{C}$ is exactly equal to $\mathbf{c^{avg}}$, while all the other compositions cannot produce the same result. This example shows that the two colluders can be uniquely identified using the designed BIBD AND-ACC codebook.

\section{Evaluation}  \label{eval}
We evaluate the performance of \sys{} on MNIST~\cite{lecun1998mnist} and CIFAR10~\cite{krizhevsky2009learning} datasets and two different neural network architectures: convolutional neural networks and wide residual networks. The topologies of these two models are summarized in Table~\ref{tab:bench}. The fingerprints are embedded in the first convolutional layer of the underlying neural network. Since orthogonal fingerprinting can be considered as a sub-category of coded fingerprinting, we focus on the comprehensive evaluation of latter one. Both MNIST-CNN and CIFAR10-WRN benchmarks are used to assess the performance of coded fingerprinting while only MNIST-CNN benchmark is used to demonstrate the workflow of orthogonal fingerprinting. 

\begin{table*}[t]
\centering
\caption{Benchmark neural network architectures. Here, ${64C3(1)}$ indicates a convolutional layer with $64$ output channels and ${3\times3}$ filters applied with a stride of 2, ${MP2(1)}$ denotes a max-pooling layer over regions of size ${2\times2}$ and stride of 1, and ${512FC}$ is a fully-connected layer consisting of $512$ output neurons. ReLU is used as the activation function in all the two benchmarks.}
\label{tab:bench}
\begin{tabular}{|c|c|c|}
\hline
Dataset & Model Type & Architecture    \\ \hline
MNIST   & CNN        & 784-32C3(1)-32C3(1)-MP2(1)-64C3(1)-64C3(1)-512FC-10FC \\ \hline
CIFAR10 & WRN & Please refer to \cite{zagoruyko2016wide}            \\ \hline
\end{tabular}
\end{table*}

\subsection{Coded Fingerprinting Evaluation}  \label{coded_eval}
In the evaluations of coded fingerprinting, we use a $(31,6,1)$-BIBD AND-ACC codebook ($\mathbf{C}$) and assign each column as a code-vector for individual users. The codebook can accommodate $ n =\frac{v(v-1)}{k(k-1)} = 31$ users and is resilient to at most $(k-1)=5$ colluders, theoretically. The embedding strength in Equation~\ref{eq:embed_loss} is set to $\gamma=0.1$ and the pre-trained host neural network is fine-tuned with the additional embedding loss for $20$ epochs in order to embed the fingerprints. The threshold for extracting the code-vector is set to $\tau = 0.85$ in all experiments. We perform a comprehensive examination of the \sys{}' performance in the rest of this paper.

\vspace{0.5em}
\noindent\textbf{Fidelity.} To show that the insertion of fingerprints does not impair the original task, we compare the test accuracy of the baseline (host neural network without fine-tuning), the fine-tuned model without embedding fingerprints, and the fine-tuned model with fingerprints embedded. The comparison of results are summarized in Table~\ref{tab:test_acc_comparison}. It can be observed from the table that embedding fingerprints in the (deep) neural network does not induce any accuracy drop and can even slightly improve the accuracy of the fine-tuned model. Thus, \sys{} meets the fidelity requirement listed in Table~\ref{tab:required}. 

\begin{table*}[]
\centering
\caption{Fidelity requirement. The baseline accuracy is preserved after fingerprint embedding in the underlying benchmarks.}
\label{tab:test_acc_comparison}
\begin{tabular}{|c|c|c|c|c|c|c|}
\hline
Benchmark          & \multicolumn{3}{c|}{MNIST-CNN}                                               & \multicolumn{3}{c|}{CIFAR10-WRN}                                         \\ \hline
Setting            & Baseline & \begin{tabular}[c]{@{}c@{}}Fine-tune without \\ fingerprint\end{tabular} & \begin{tabular}[c]{@{}c@{}}Fine-tune with \\ fingerprint\end{tabular} & Baseline & \begin{tabular}[c]{@{}c@{}}Fine-tune without \\ fingerprint\end{tabular} & \begin{tabular}[c]{@{}c@{}}Fine-tune with\\  fingerprint\end{tabular} \\ \hline
Test Accuracy (\%) & 99.52    & 99.66             & 99.72       & 91.85    & 91.99                         & 92.03         \\ \hline
\end{tabular}
\end{table*}

\vspace{0.5em}
\noindent{\textbf{Uniqueness.}} The uniqueness of code modulated fingerprinting originates from the $(v,k.1)$-BIBD ACC codebook. Since the code-vector assigned to each user is the bit complement of columns of the incidence matrix~\cite{wu2004collusion} which has no repeated columns, individual users are uniquely identified by the associated ACC code-vectors.

\vspace{0.5em}
\noindent {\textbf{Scalability.}} Due to the intrinsic requirement of model distribution and sharing, the fingerprinting methodology should be capable of accommodating a large number of users. For a $(v,k,1)$-BIBD ACC codebook, the maximum number of users is decided by the code length $v$ and the block size $k$:
 \begin{equation*}
 	 n=\frac{v(v-1)}{k(k-1)}.
 \end{equation*}
 
Systematic approaches to design various families of BIBDs have been well studied in previous literature~\cite{colbourn2006handbook,rodger2008design}. For instance, Steiner triple systems are families of $(v,3,1)$-BIBD systems and are shown to exist if and only if $v \equiv 1$ or $3$ (mod 6)~\cite{rodger2008design}. An alternative method to design BIBDs is to use projective and affine geometry in $d-$dimension over $Z_p$, where $p$ is of prime power. $(\frac{p^{d+1} -1}{p-1}, p+1, 1)$-BIBDs and $(p^d, p, 1)$-BIBDs can be constructed from projective and affine geometry~\cite{colbourn2006handbook, lidl1994introduction}. By choosing a large dimension of fingerprints in Steiner triple systems, or using projective geometry in a high dimensional space, the number of users allowed in our proposed framework can be sufficiently large. Therefore, the scalability of \sys{} is guaranteed by a properly designed BIBD ACC codebook. By expanding the ACC codebook, \sys{} supports IP protection and DRM when new users join in the model distribution system.

\vspace{0.5em}
\noindent{\textbf{Robustness, Reliability, and Integrity.}} We evaluate the robustness of \sys{} against fingerprints collusion attack and model modifications, including parameter pruning as well as model fine-tuning on MNIST and CIFAR10 benchmarks. For all attacks, we assume the fingerprinting method as well as the positions of the marked layers are known to the attackers. The code-vector extraction and colluder detection scheme are described in Section~\ref{coded_detect}. 

We use a $(31,6,1)$-BIBD AND-ACC codebook and assume there are $31$ users in total. For a given number of colluders, $10,000$ random simulations are performed to generate different colluders sets from all users. When the colluder set is too large to be uniquely identified by the BIBD AND-ACC codebook, we consider all feasible colluder sets that match the extracted code-vector resulting from the fingerprints collusion and take the mean value of the detection rates as well as the false alarm rates. The average performance over $10,000$ random tests is used as the final metric. The details of the robustness tests against the three aforementioned attacks are explained in the following sections.

\begin{figure}[ht!]
\centering
\includegraphics[width=0.4\textwidth]{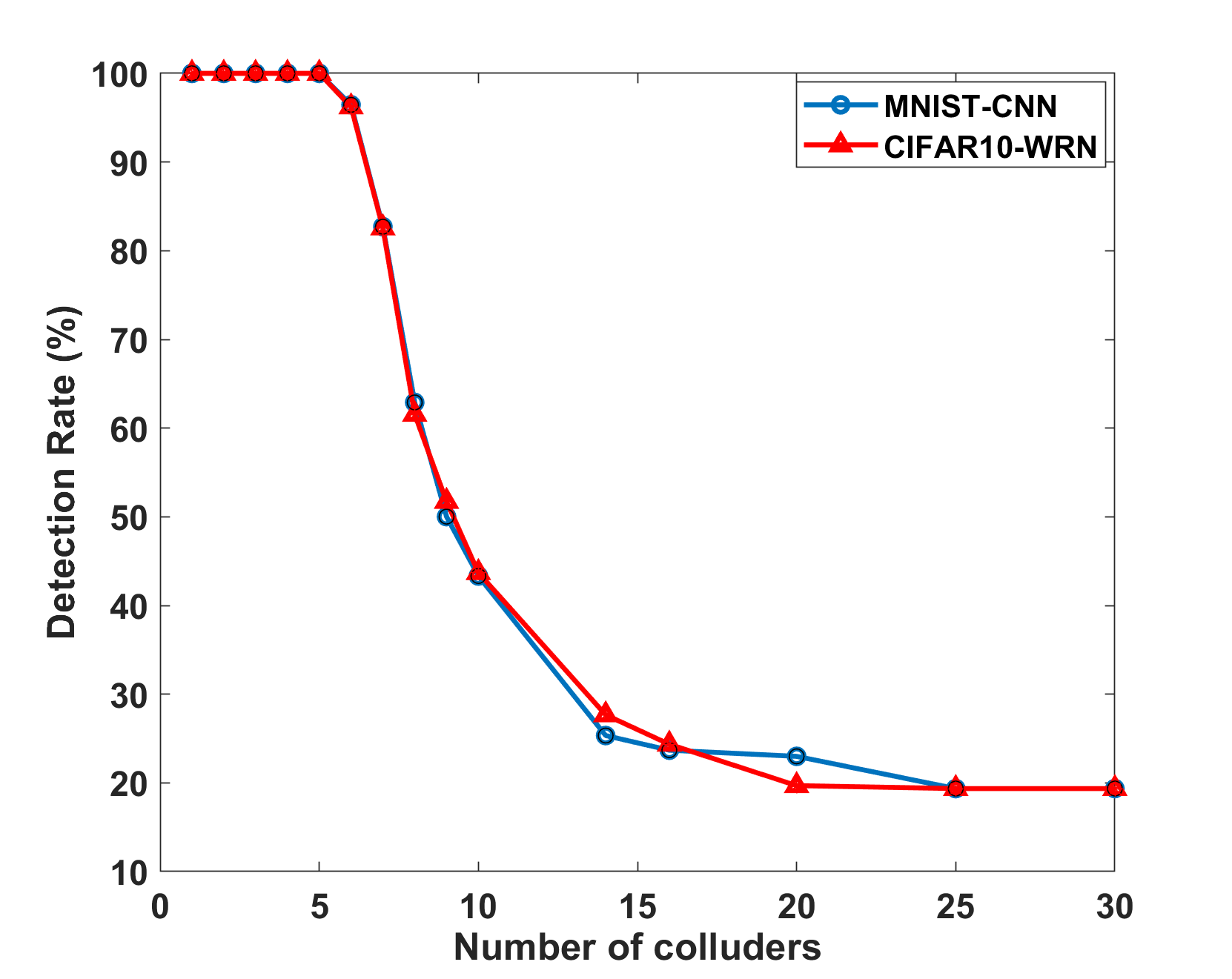} 
\caption{\label{fig:v31_detection} Detection (true positive) rates of fingerprints averaging attack. Using $(31,6,1)$-BIBD ACC codebook, up to 5 colluders can be uniquely identified with $100\%$ detection rate. }
\vspace{-0.3em}
\end{figure}

\vspace{0.5em}
\noindent \textbf{(I) Fingerprints collusion.} Figure~\ref{fig:v31_detection} shows the detection rates of \sys{} when different number of users participate in the collusion attack. As can be seen from Figure~\ref{fig:v31_detection}, the detection rate is $100\%$ when the number of colluders is smaller or equal to $5$, which means the collusion resilience level is $K_{max} = 5$ with the $(31,6,1)$-BIBD ACC codebook. When the number of colluders further increases, the detection rate starts to decrease, and finally reaches a stable value at $19.35\%$.  

\begin{figure}[ht!]
\centering
\vspace{-0.65em}
\includegraphics[width=0.4\textwidth]{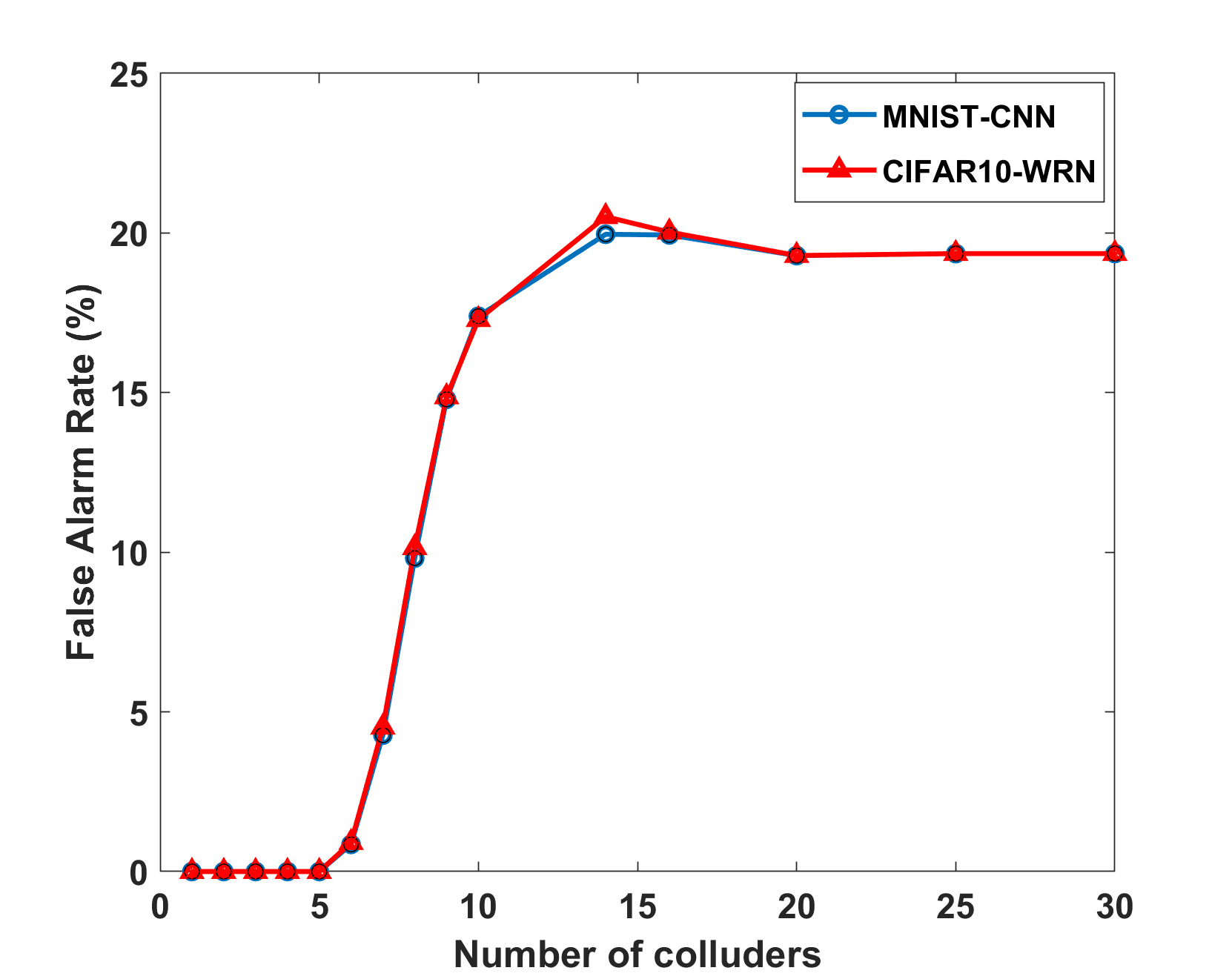} 
\caption{\label{fig:v31_falseAlarm} False alarm (false positive) rates of fingerprints averaging attack. Using a $(31,6,1)$-BIBD ACC codebook, no false accusement occurs if the number of colluders is smaller or equal to 5. }
\end{figure}

Along with the evaluation of detection rates, we also assess the false alarm rates of \sys{} using $10,000$ random simulations and summarize the results in Figure~\ref{fig:v31_falseAlarm}. It can be seen that the false accuse rate remains $0\%$ when the number of colluders does not exceed $5$, which is consistent with $K_{max}=5$ found in the evaluations of detection rates. When the number of colluders increases, the false alarm rate first increases and stays at a stable value at the end. 

Comparing the detection performance of \sys{} on MNIST-CNN and CIFAR10-WRN benchmarks shown in Figures~\ref{fig:v31_detection}~and~\ref{fig:v31_falseAlarm}, one can observe that the detection rates and the false alarm rates are approximately the same for two benchmarks given the same number of colluders. The consistency across benchmarks derives from the correct code-vector extraction and the unique identification property of BIBD ACC codebooks. The colluder detection scheme of \sys{} can be considered as a high-level protocol which is completely independent of the underlying network architecture and the dataset. Therefore, our proposed framework meets the \textbf{generality} requirement listed in Table~\ref{tab:required}.  

The high detection rates and low false alarm rates corroborate that \sys{} satisfies the \textbf{reliability} and \textbf{integrity} requirements in Table~\ref{tab:required}, respectively. Furthermore, the maximal number of colluders that the system can identify with $100\%$ detection rate and $0\%$ false alarm rate is found to be $K_{max}=5$, which is consistent with the theoretical tolerance ($k-1$) given by the BIBD AND-ACC. The consistency helps the owner to choose the proper ACC codebook based on her desired collusion resilience requirement. 

\begin{figure}[ht!]
    \centering
    \begin{subfigure}[h]{0.45\columnwidth}
        \centering
        \includegraphics[width=0.98\columnwidth]{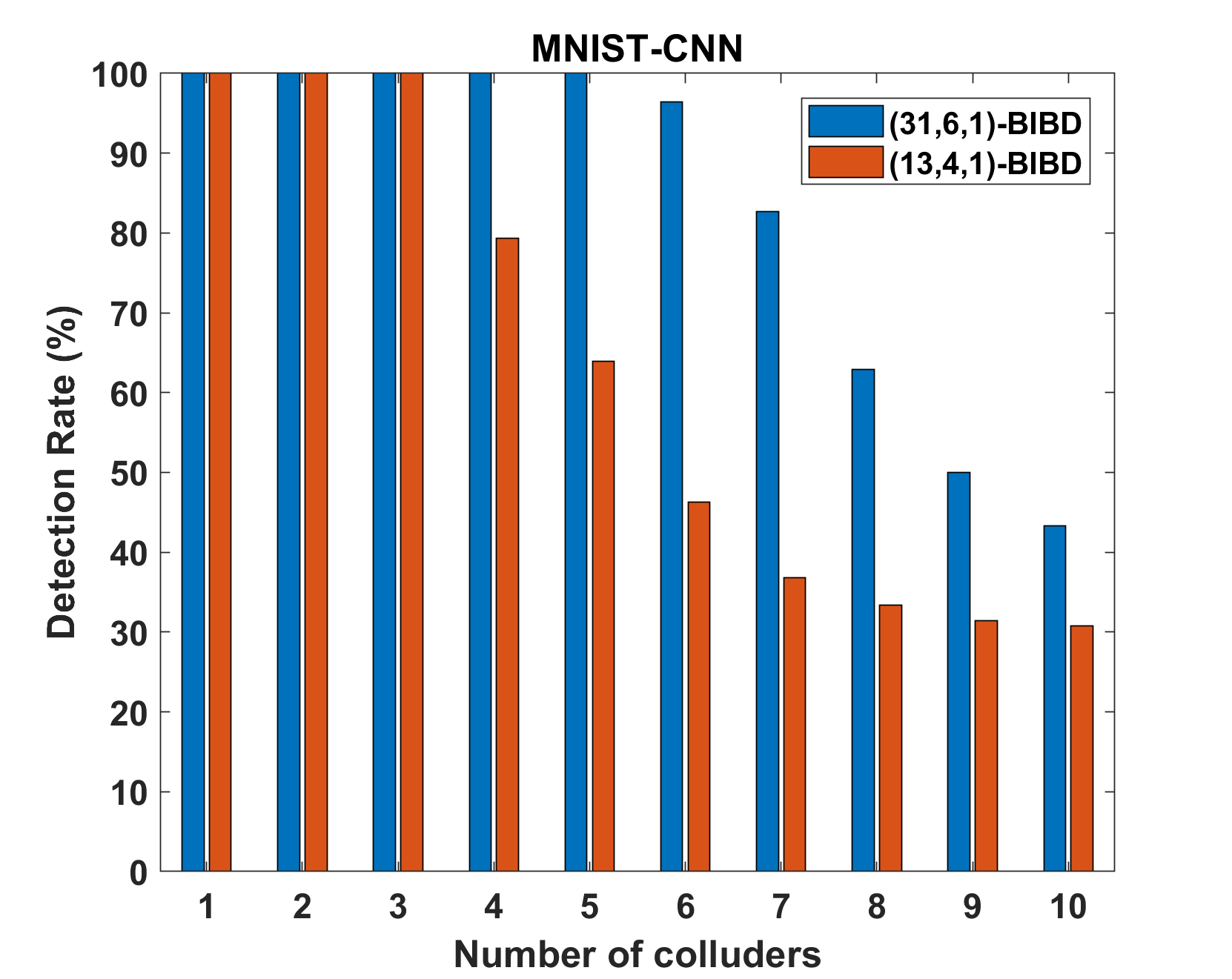} 
        \caption{{\label{fig:codebook_mnist_detect}}}
    \end{subfigure}
    ~ 
    \begin{subfigure}[h]{0.45\columnwidth}
        \centering
        \includegraphics[width=0.98\columnwidth]{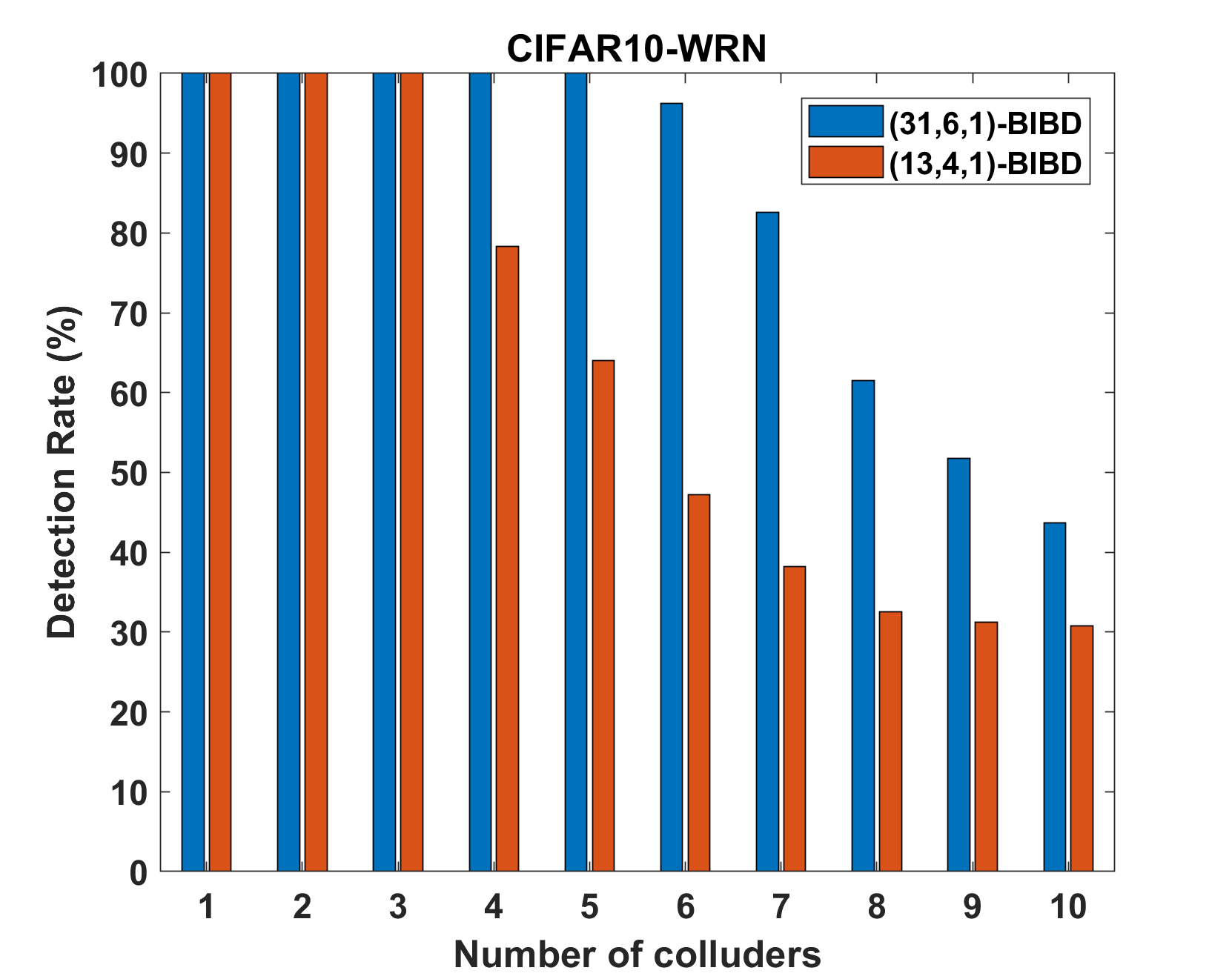}
        \caption{\label{fig:codebook_cifar10_detect}}
    \end{subfigure}
    \caption{Detection rates of fingerprints collusion attacks on (a) MNIST-CNN and (b) CIFAR10-WRN benchmarks when different ACC codebooks are used. The codebook with larger block size $k$ has better detection capability of collusion attacks.}
\end{figure}

\begin{figure}[ht!]
    \centering
    \begin{subfigure}[h]{0.45\columnwidth}
        \centering
        \includegraphics[width=0.98\columnwidth]{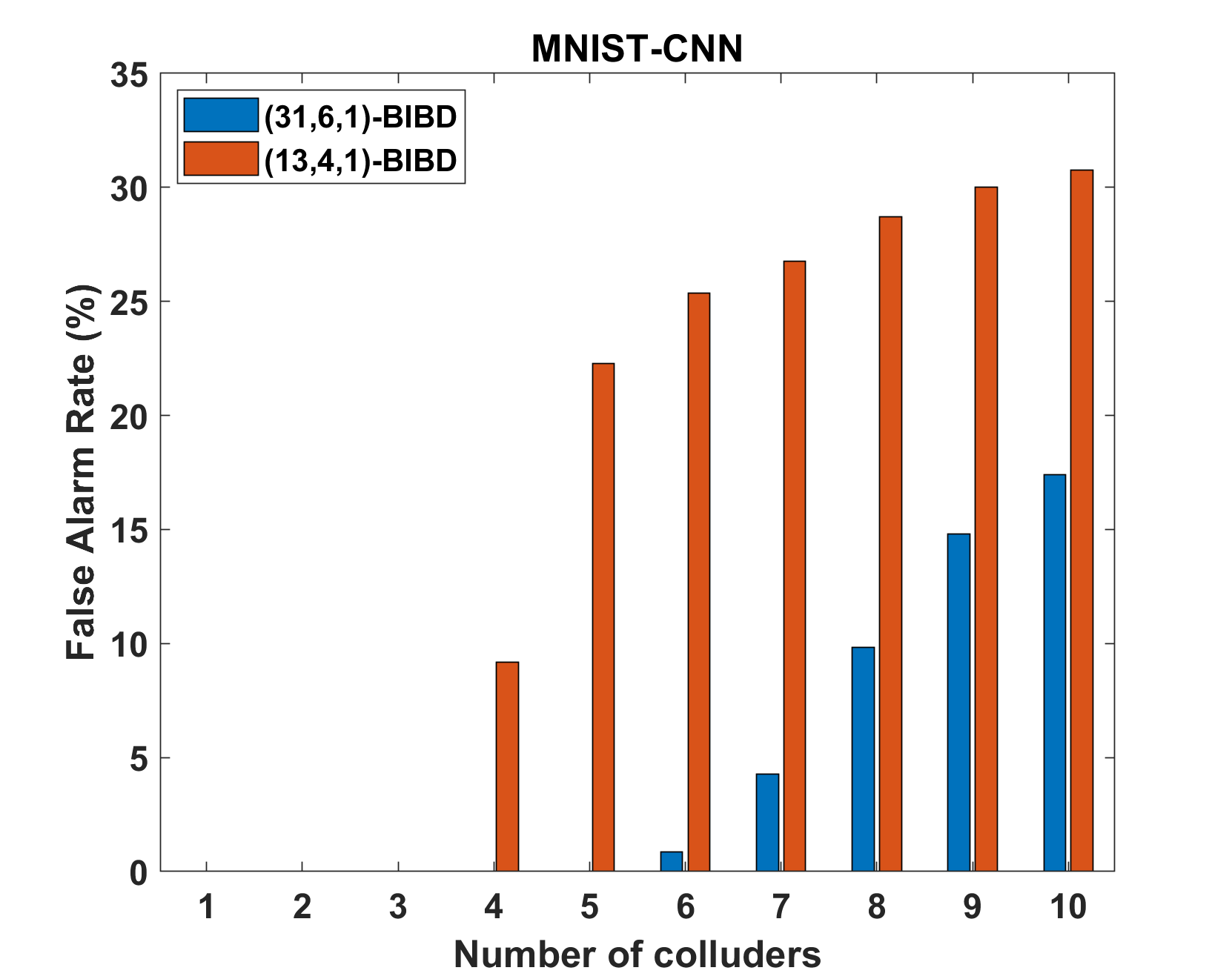} 
        \caption{{\label{fig:codebook_mnist_falseAlarm}}}
    \end{subfigure}
    ~ 
    \begin{subfigure}[h]{0.45\columnwidth}
        \centering
        \includegraphics[width=0.98\columnwidth]{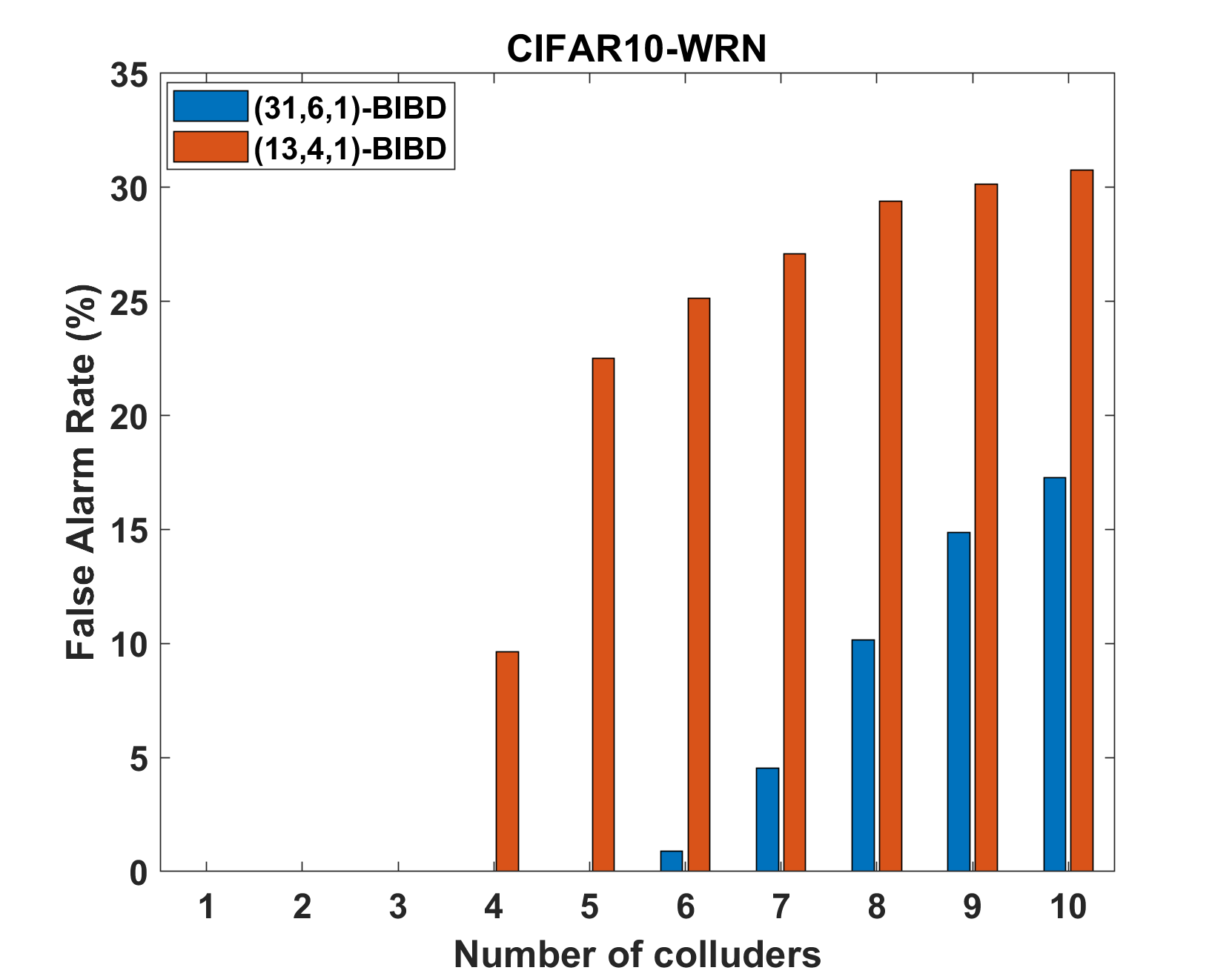}
        \caption{\label{fig:codebook_cifar10_falseAlarm}}
    \end{subfigure}
    \caption{False alarm rates of fingerprints collusion attacks on (a) MNIST-CNN and (b) CIFAR10-WRN benchmarks. Given the same number of colluders, the $(31,6,1)$-BIBD ACC codebook has lower false alarm rates than the $(13.4.1)$-BIBD ACC codebook.}
\end{figure}

For a comprehensive evaluation of \sys{}, we further compare the robustness of our proposed framework when the $(31,6,1)$-BIBD ACC and the $(13,4,1)$-BIBD ACC codebook are used. The detection rates of the fingerprints collusion attacks on MNIST and CIFAR10 datasets are shown in Figures~\ref{fig:codebook_mnist_detect}~and~\ref{fig:codebook_cifar10_detect}, respectively. The false alarm rates are shown in Figures~\ref{fig:codebook_mnist_falseAlarm}~and~\ref{fig:codebook_cifar10_falseAlarm}. The comparison between two codebooks shows how the design of BIBD-ACC codebooks affects the collusion resistance of \sys{}. Particularly, it can be observed that the $(31,6,1)$-BIBD AND-ACC codebook has a collusion resilience level $K_{max}=5$ while the $(13,4,1)$-BIBD AND-ACC only has the resilience level $K_{max}=3$. In addition, the $(31,6,1)$-BIBD codebook has higher detection rate as well as lower false accuse rate compared to $(13,4,1)$-BIBD codebook given a specific number of colluders. Same conclusions hold for the collusion resilience against parameter pruning and model fine-tuning attacks. For simplicity, the results are not presented here.

\begin{figure}[ht!]
\centering
\includegraphics[width=0.4\textwidth]{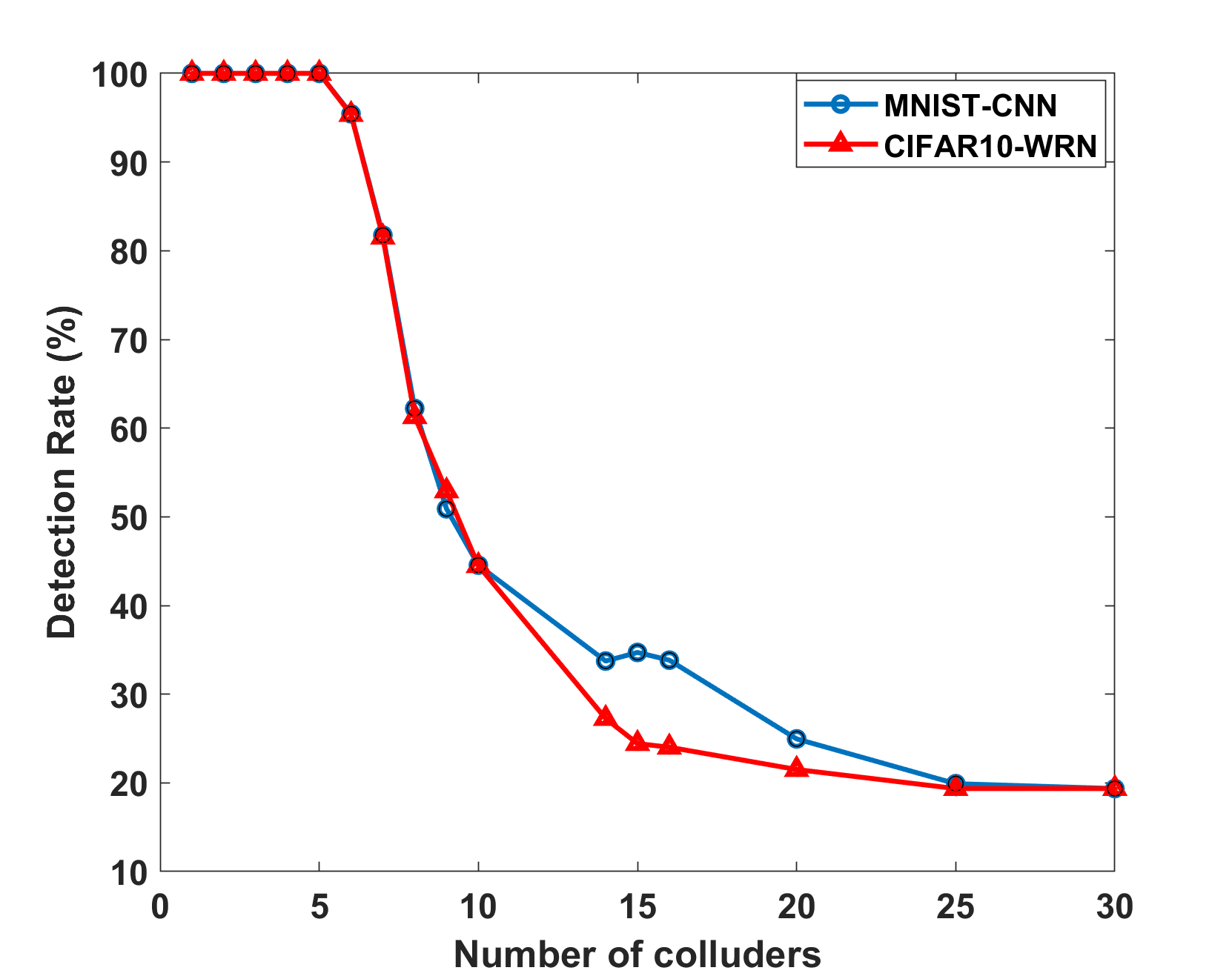} 
\caption{\label{fig:v31_finetune_detect} Detect rates of fingerprint collusion with model fine-tuning. \sys{} attains high detection rate and the same resilience level $K_{max}=5$ even if the marked neural network is fine-tuned. }
\end{figure}

\begin{figure}[ht!]
\centering
\includegraphics[width=0.4\textwidth]{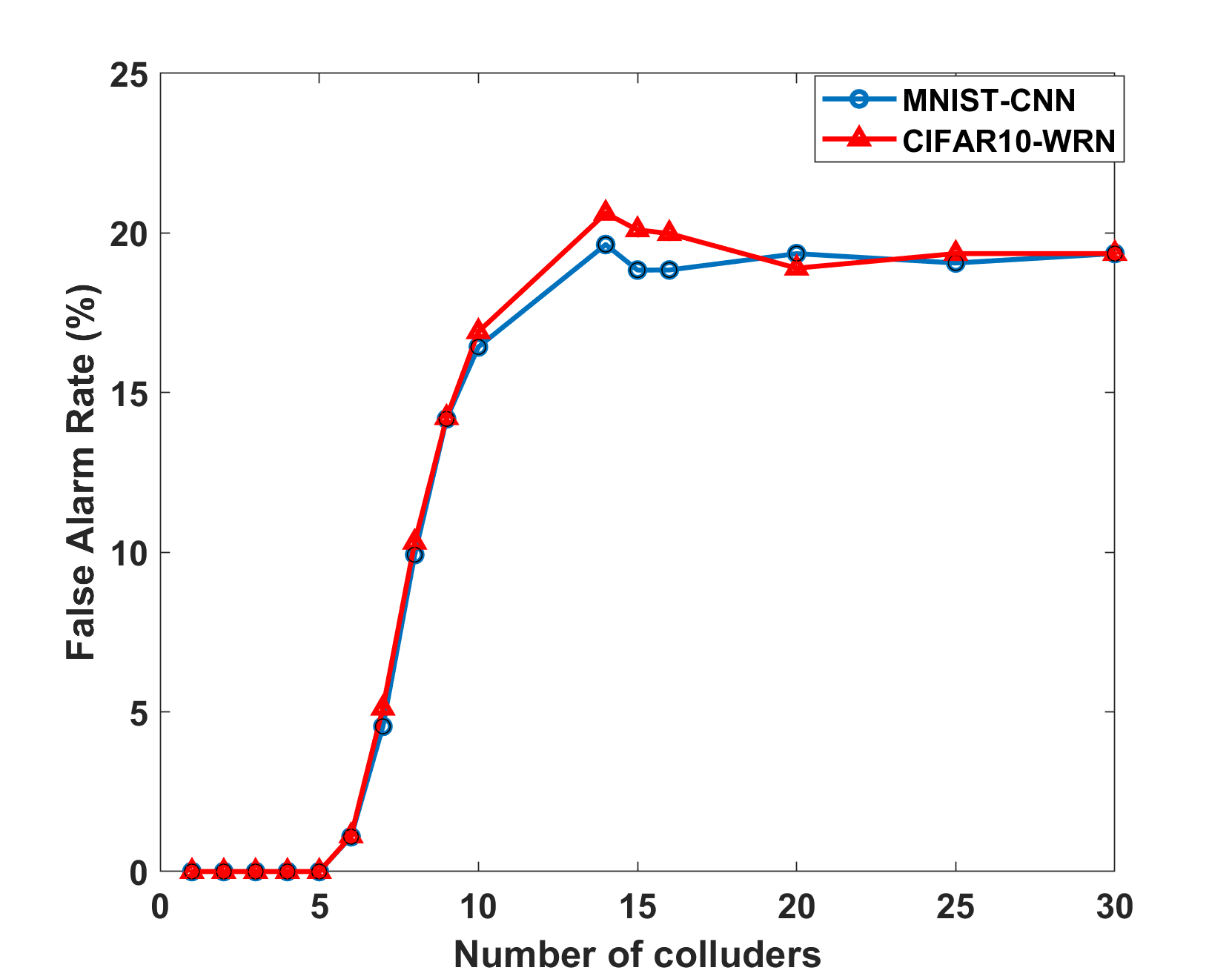} 
\caption{\label{fig:v31_finetune_falseAlarm} False alarm rates of fingerprint collusion with model fine-tuning. The collusion resilience level $(K_{max}=5)$ is not affected by fine-tuning attack. }
\end{figure}

\vspace{0.5em}
\noindent \textbf{(II) Model fine-tuning.} To evaluate the robustness against the fine-tuning attack, we retrain the fingerprinted model using only conventional cross-entropy loss as the objective function. The code-vector extraction and colluder detection scheme are the same as in the evaluation of fingerprints collusion attacks. The detection rates and false alarm rates of \sys{} on MNIST and CIFAR10 datasets are shown in Figures~\ref{fig:v31_finetune_detect}~and~\ref{fig:v31_finetune_falseAlarm}, respectively. Compared with Figures~\ref{fig:v31_detection}~and~\ref{fig:v31_falseAlarm}, where the robustness against collusion attacks is evaluated without model fine-tuning, the same trend can be observed and the collusion resistance level remains the same $K_{max}=5$, showing that \sys{} is robust against model fine-tuning attacks. 

\begin{figure}[ht!]
\centering
\includegraphics[width=0.4\textwidth]{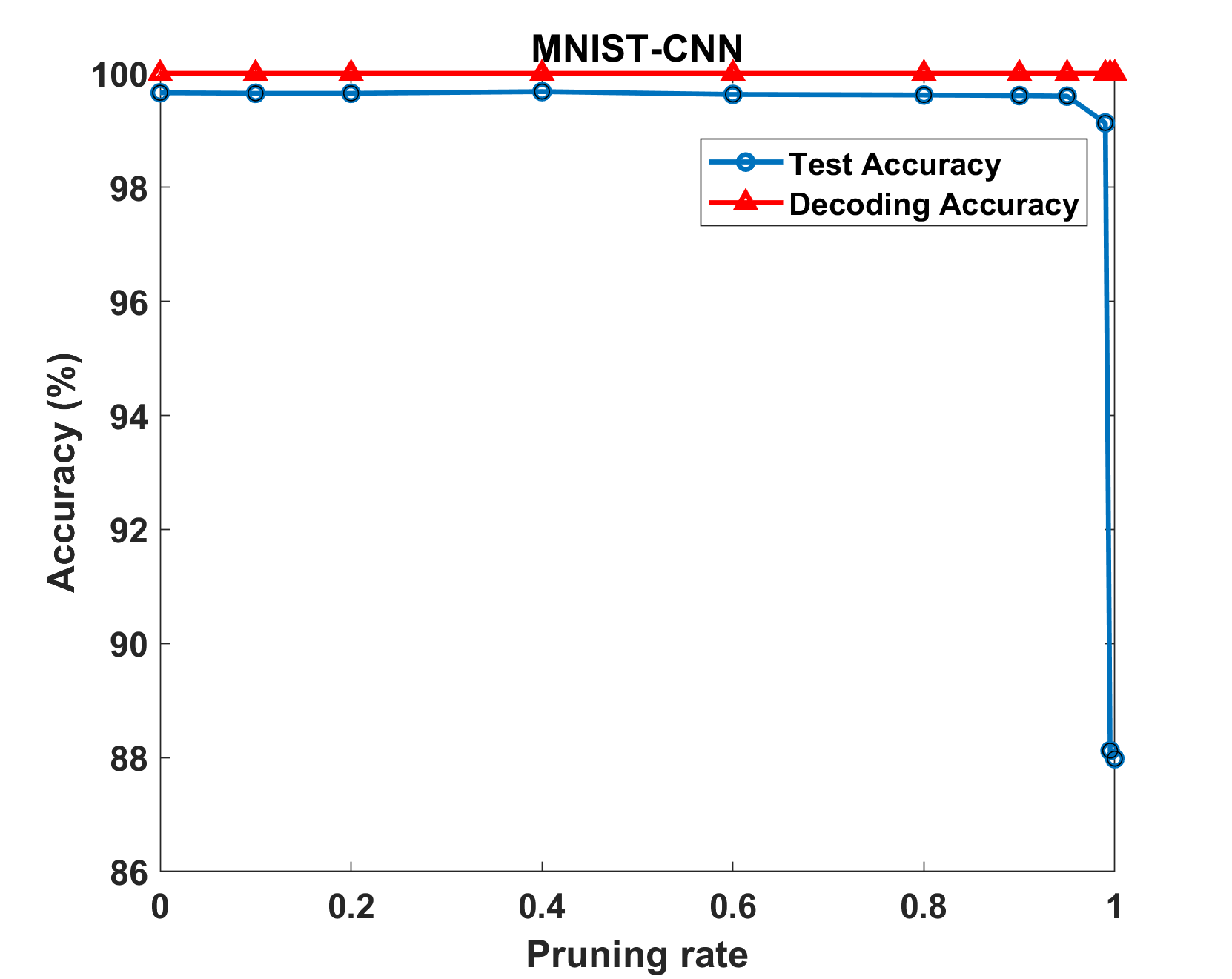} 
\caption{\label{fig:v31_prune_mnist_decode_acc} Code-vector extraction accuracy and test accuracy under different pruning rates. The test accuracy of MNIST-CNN drops when the pruning rate when the pruning rates is larger than $95\%$ while the ACC decoding accuracy remains at $100\%$. }
\end{figure}

\begin{figure}[ht!]
\centering
\includegraphics[width=0.4\textwidth]{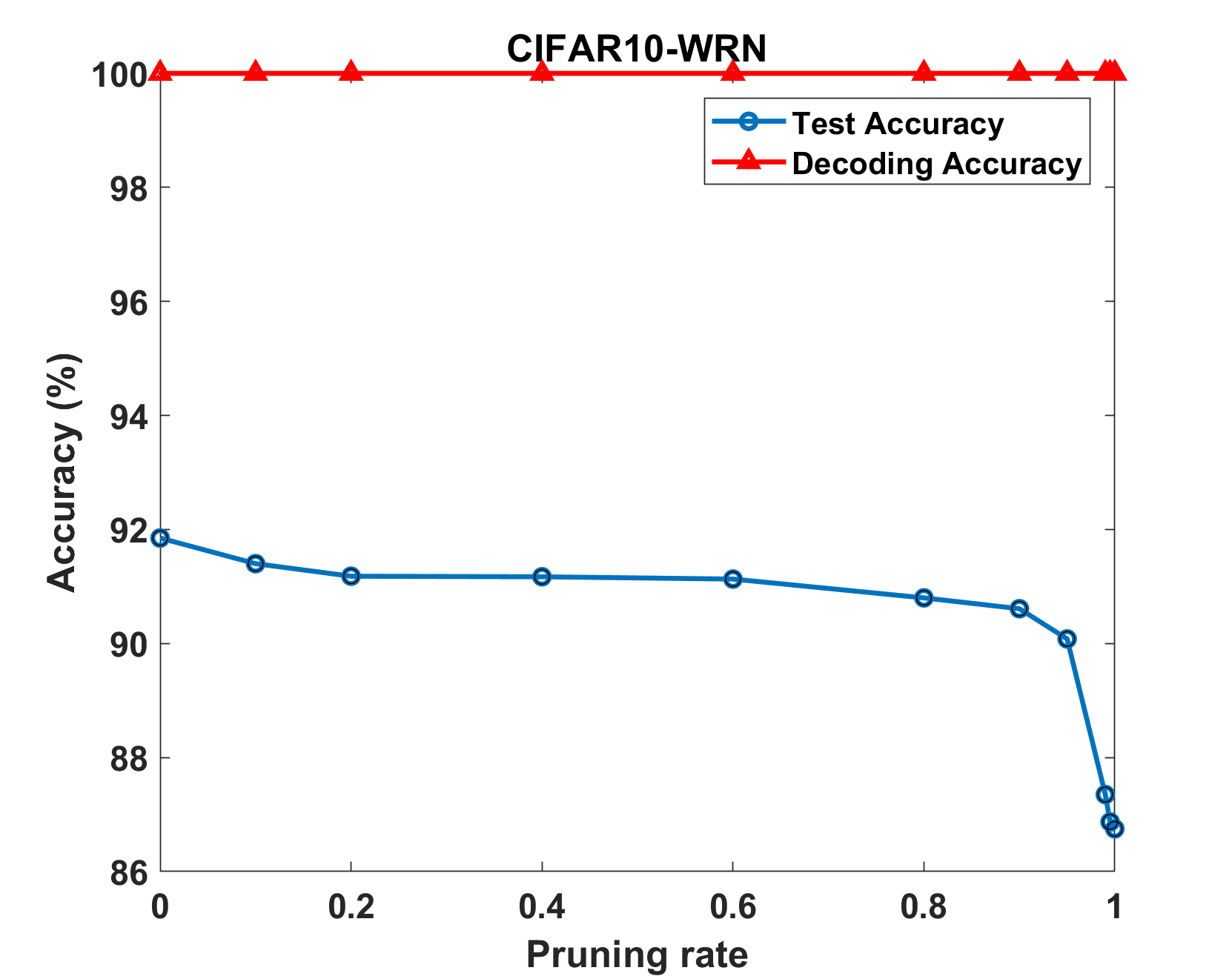} 
\caption{\label{fig:v31_prune_cifar10_decode_acc} Code-vector extraction accuracy and test accuracy under different pruning rates. The ACC decoding accuracy is robust up to $99.99\%$ pruning rate while the test accuracy of CIFAR10-WRN benchmark degrades when the pruning rate is greater than $90\%$. }
\end{figure}

\vspace{0.5em}
\noindent \textbf{(III) Parameter pruning.} Parameter pruning alters the weights of the marked neural network. As such, we first evaluate the code-vector extraction (decoding) accuracy of \sys{} under different pruning rates. Figures~\ref{fig:v31_prune_mnist_decode_acc}~and~\ref{fig:v31_prune_cifar10_decode_acc} show the results on MNIST and CIFAR10 datasets, respectively. One can see that increasing the pruning rate leads to the drop of the test accuracy, while the code-vector can always be correctly decoded with $100\%$ accuracy. The superior decoding accuracy of  the AND-ACC code-vectors under various pruning rates corroborates the robustness of our designed fingerprint code-vectors against the parameter pruning attack.

\begin{figure}[ht!]
\centering
\includegraphics[width=0.4\textwidth]{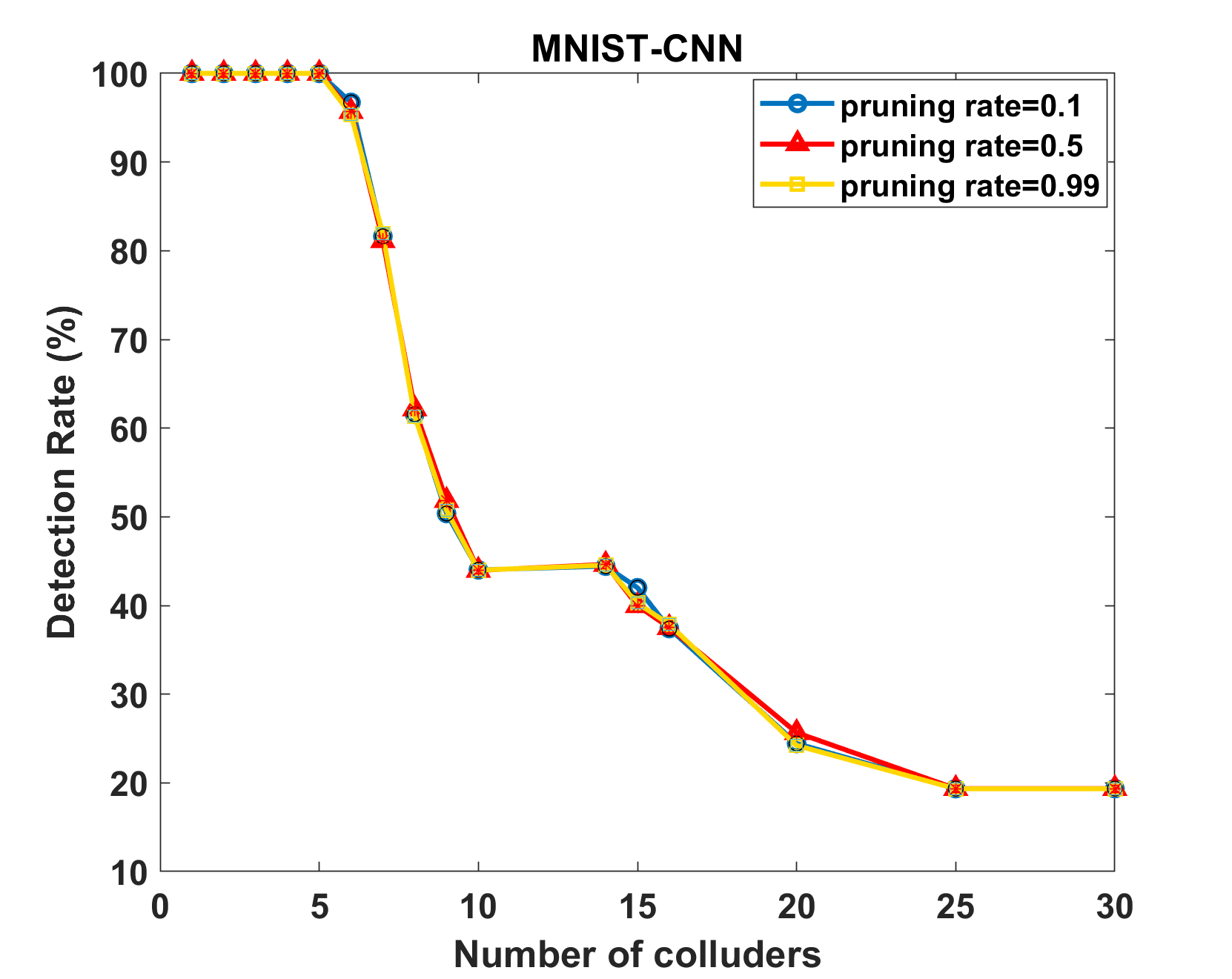} 
\caption{\label{fig:v31_mnist_prune_detect} Detection rates of fingerprint collusion with three different pruning rates on MNIST dataset. The collusion resistance level  $K_{max}=5$ is robust up to $99\%$ parameter pruning. }
\end{figure}

\begin{figure}[ht!]
\centering
\includegraphics[width=0.4\textwidth]{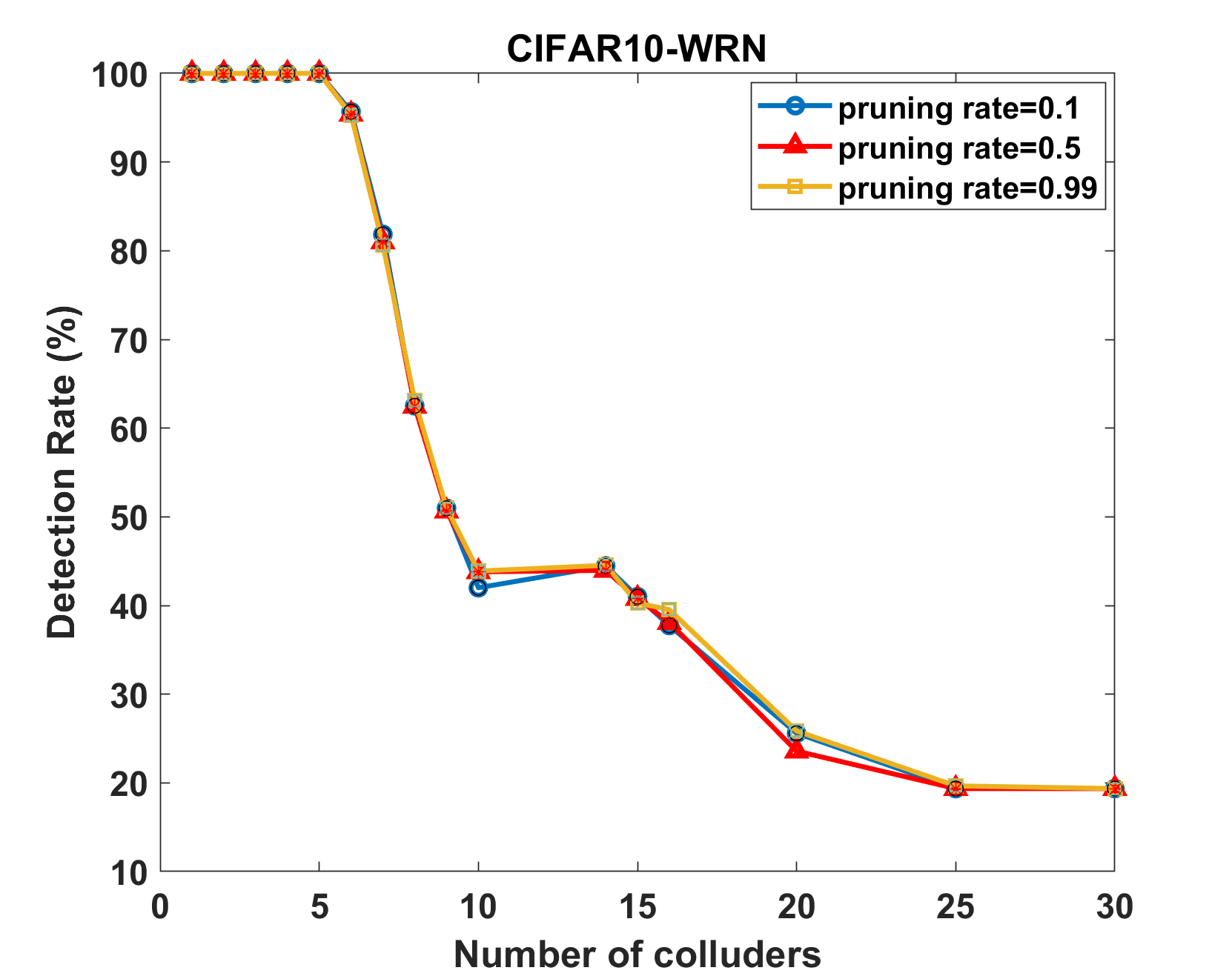} 
\caption{\label{fig:v31_cifar10_prune_detect} Detection rates of the fingerprints collusion attack with three different pruning rates on CIFAR10 dataset. The collusion resistance level remains at $K_{max}=5$ when the parameter pruning is mounted on fingerprints collusion attacks.}
\end{figure}

We further assess the robustness of \sys{} for colluders identification against parameter pruning. Figures~\ref{fig:v31_mnist_prune_detect}~and~\ref{fig:v31_cifar10_prune_detect} show the detection rates of \sys{} under three different pruning rates ($10\%, 50\%, 99\%$) using MNIST-CNN and CIFAR10-WRN benchmarks, respectively. Similar to the trend shown in Figure~\ref{fig:v31_detection}, the same collusion resilience level $K_{max}=5$ is observed in Figures~\ref{fig:v31_mnist_prune_detect}~and~\ref{fig:v31_cifar10_prune_detect}; suggesting that the \textbf{reliability} and the \textbf{robustness} criteria are satisfied for the parameter pruning attacks as well.

\begin{figure}[ht!]
\centering
\includegraphics[width=0.4\textwidth]{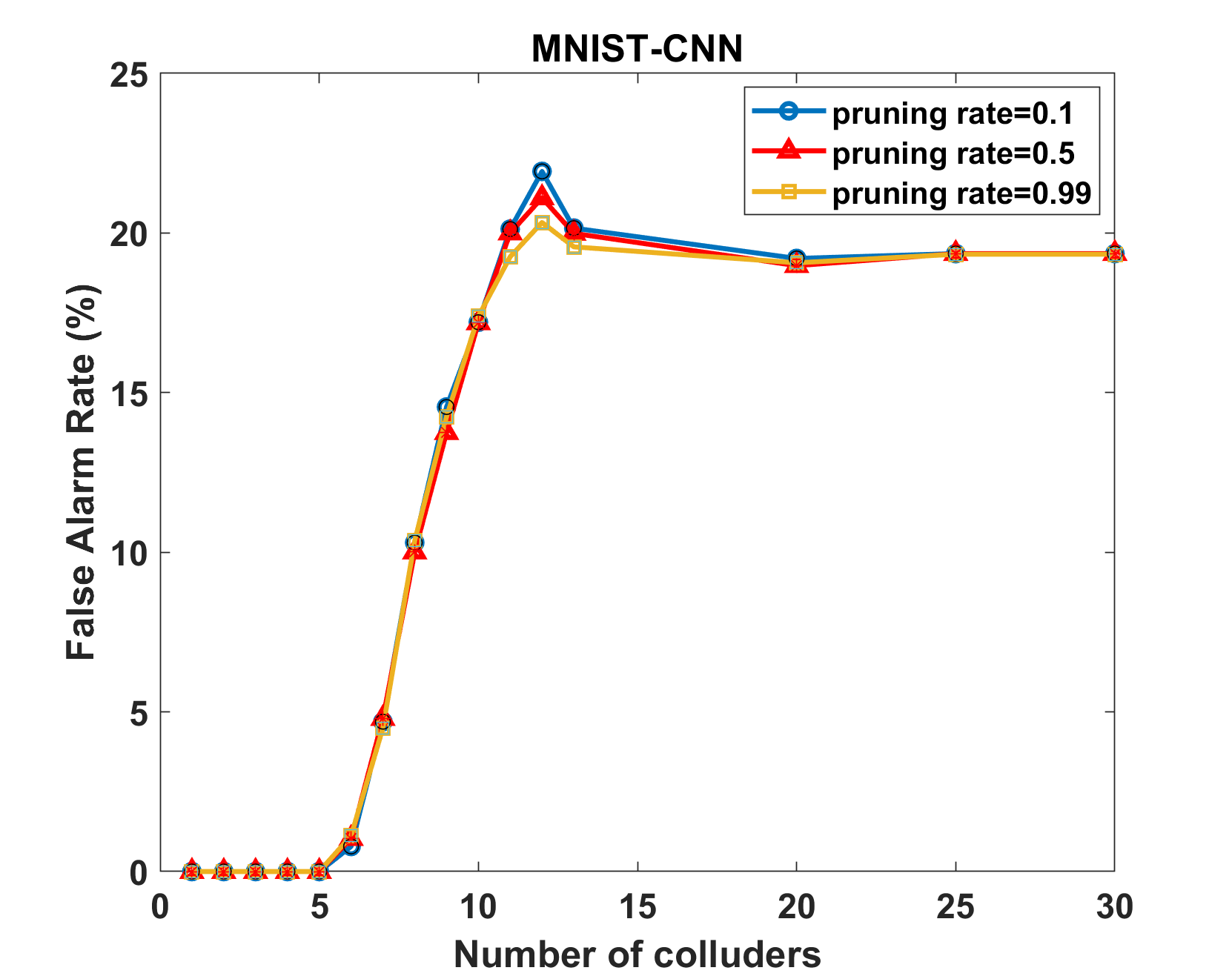} 
\caption{\label{fig:v31_mnist_prune_falseAlarm} False alarm rates of fingerprint collusion attacks on MNIST dataset where three pruning rates are tested. No false alarms will occur if the number of colluders does not exceed 5, showing the robustness and integrity of \sys{} against parameter pruning.}
\end{figure}

\begin{figure}[ht!]
\centering
\includegraphics[width=0.4\textwidth]{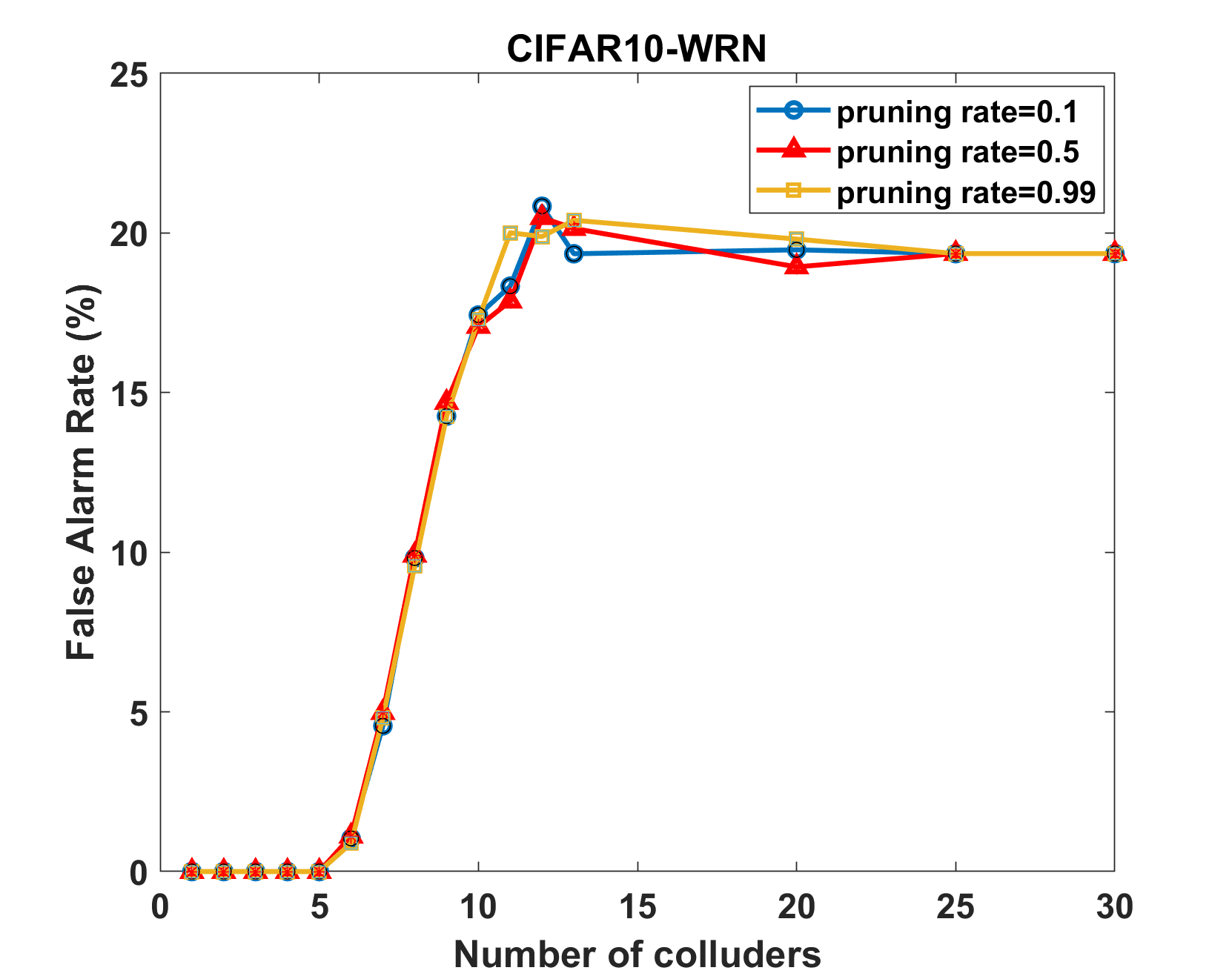} 
\caption{\label{fig:v31_cifar10_prune_falseAlarm} False alarm rates of collusion attacks on CIFAR10 dataset where three pruning rates are used. Innocent users will not be incorrectly accused if the number of colluders is at most 5. }
\vspace{-0.5em}
\end{figure}

To assess the integrity of \sys{}, the false alarm rates under three different pruning rates are also evaluated on MNIST and CIFAR10 datasets. From the experimental results shown in Figures~\ref{fig:v31_mnist_prune_falseAlarm}~and~\ref{fig:v31_cifar10_prune_falseAlarm}, one can see that no false accusement will occur if the number of colluders is smaller or equal to $K_{max}=5$, which is consistent with the evaluation of fingerprints collusion attacks. This consistency indicates that our colluder detection scheme satisfies the \textbf{integrity} criterion and is robust against parameter pruning attacks. 

In conclusion, the high detection rates and low false alarm rates under different attack scenarios corroborate that \sys{} satisfies the robustness, reliability, and integrity requirements discussed in Table~\ref{tab:required}. The consistency of detection performance across benchmarks indicates that \sys{} meets the generality requirement.

\noindent{\textbf{Efficiency.}} Here, we discuss the efficiency of the fingerprinting methodology in terms of the runtime overhead for fingerprints embedding and the efficiency of the AND-ACC codebook.

\begin{figure}[]
\centering
\includegraphics[width=0.4\textwidth]{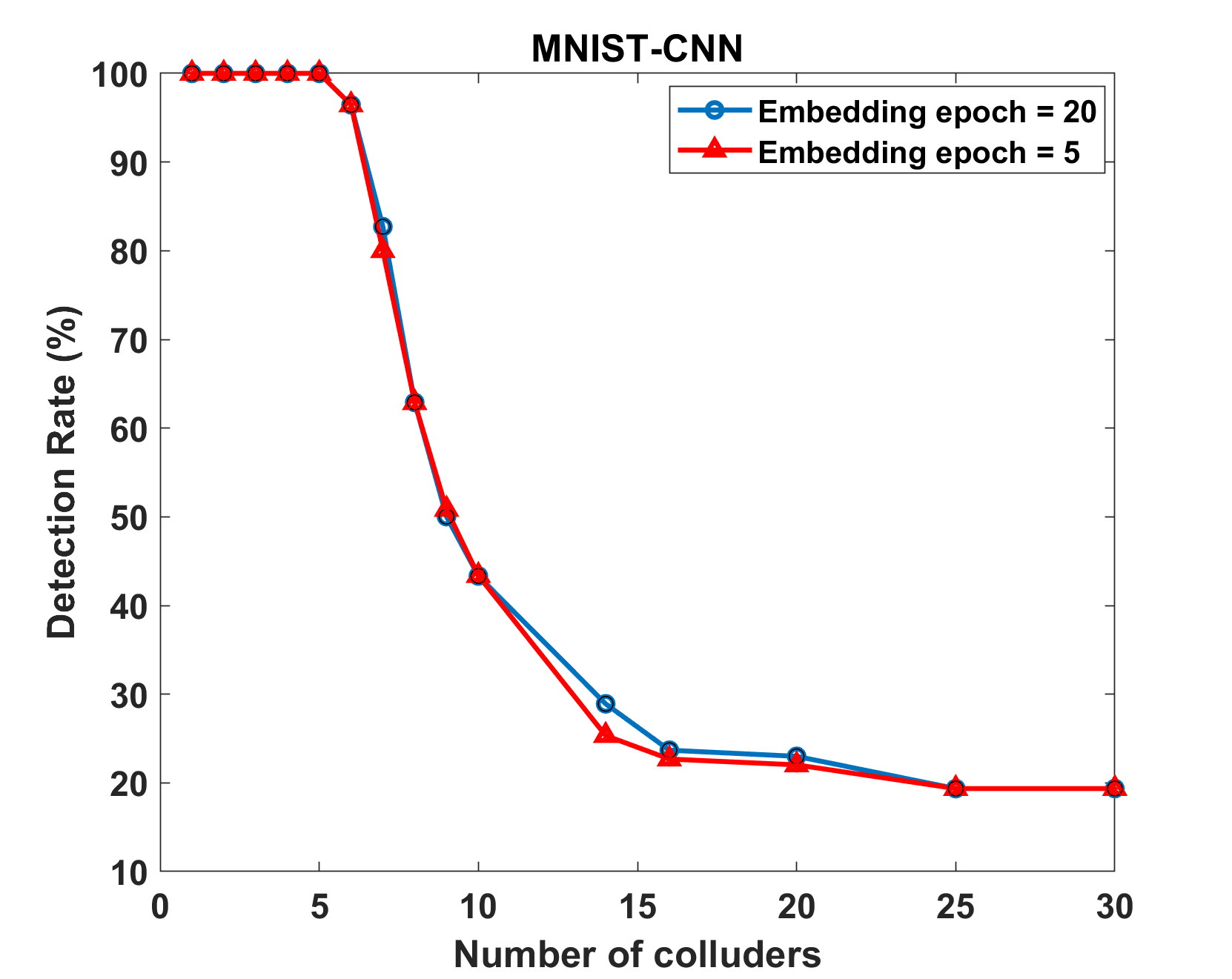} 
\caption{\label{fig:overhead_mnist_collusion_detect} Detection rates of fingerprint collusion attacks on MNIST-CNN benchmark. The same detection performance can be observed when the unmarked model is retrained for two different epochs for fingerprints embedding.}
\end{figure}

\begin{figure}[ht!]
\centering
\includegraphics[width=0.4\textwidth]{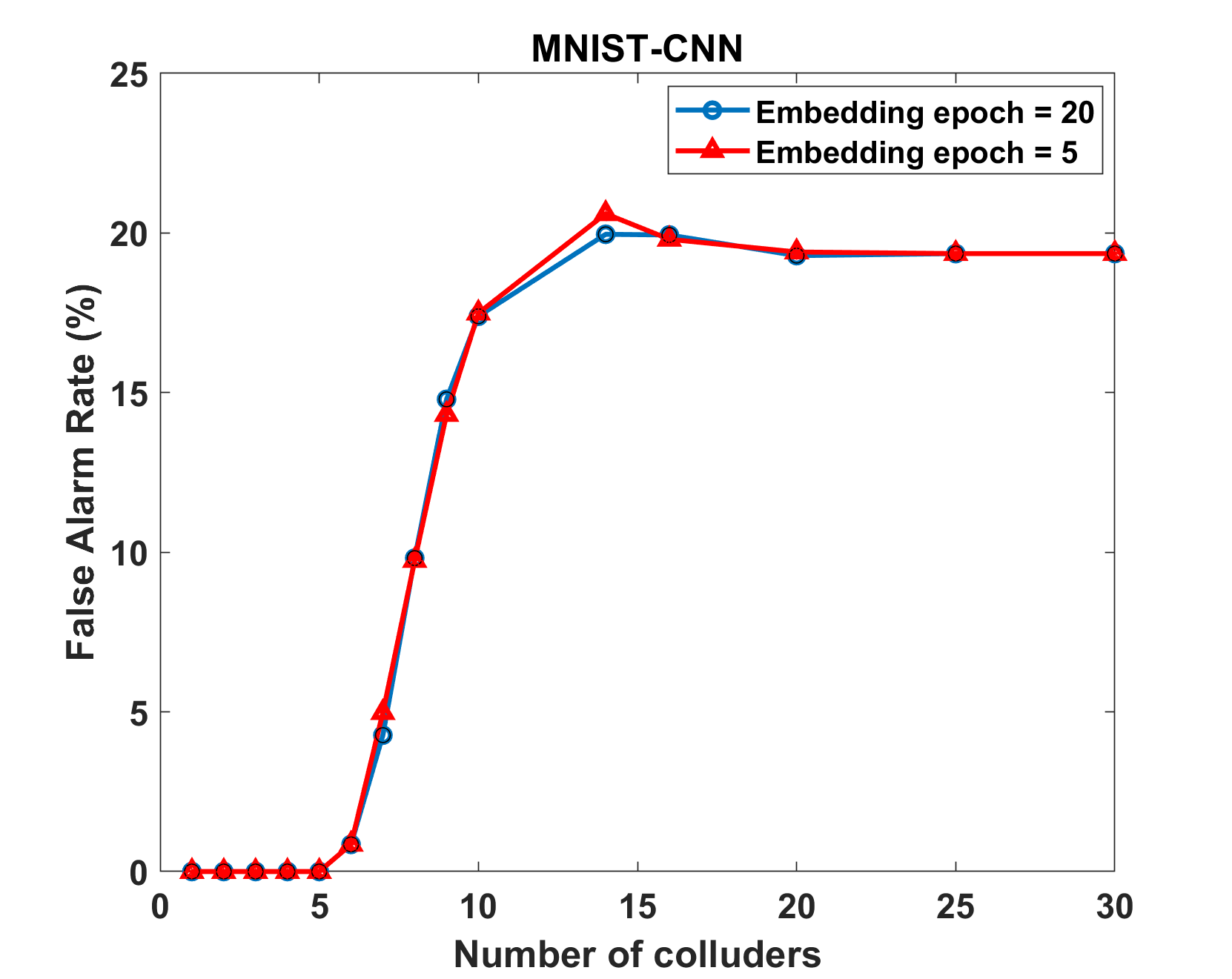} 
\caption{\label{fig:overhead_mnist_collusion_falseAlarm} False alarm rates of fingerprints collusion attacks on MNIST-CNN benchmark when two different embedding epochs are used. Retraining for 5 epochs is sufficient to ensure low false alarm rates.}
\end{figure}

\begin{itemize}
\item  \textit{Fingerprints Embedding Overhead.}
Since each distributed model needs to be retrained with the fingerprint embedding loss, it is necessary that the fingerprinting methodology has low runtime overhead of generating individual fingerprints. We evaluate the fingerprints embedding efficiency of \sys{} by retraining the unmarked host neural network for 20 epochs and 5 epochs. The robustness of the resulting two marked models against fingerprints collusion attacks is compared. Figures~\ref{fig:overhead_mnist_collusion_detect}~and~\ref{fig:overhead_mnist_collusion_falseAlarm} demonstrate the detection rates and false alarm rates of these two marked models using MNIST-CNN benchmark. As can be seen from the comparison, embedding fingerprints by retraining the neural network for 5 epochs is sufficient to ensure the collusion resistance of the embedded fingerprints, suggesting that the runtime overhead induced by \sys{} is negligible. We also observe that the marked models retrained for 5 epochs and 20 epochs have the same collusion resistance level against parameter pruning and model fine-tuning attacks. 

\vspace{0.5em}
\item \textit{ACC Codebook Efficiency.} In the multi-media domain, the efficiency of an AND-ACC codebook for a given collusion resistance is defined as the number of users that can be supported per basis vector: $\beta = \frac{n}{v}$. For a $(v,k,1)$-BIBD AND-ACC, the codebook efficiency is:
\begin{equation} \label{eq:code_effic}
	\beta= \frac{v-1}{k(k-1)}.
\end{equation}
Thus, for a fixed resilience level $k$, the
efficiency of an AND-ACC codebook constructed from BIBDs improves as
the code length increases~\cite{trappe2003anti}. \sys{} allows the owner to design an efficient coded fingerprinting methodology by choosing appropriate parameters for the BIBD ACC codebook.

In contrast with orthogonal fingerprinting where the number of users is the same as the fingerprint dimension  $n=v$ (thus $\beta=1$), it has been proven that a $(v,k,\lambda)$-BIBD has $n \geq v$~\cite{dinitz1992contemporary}, meaning that the codebook efficiency of the BIBD construction satisfies $\beta \geq 1$. Equation~\ref{eq:code_effic} shows the trade-off between the code length $v$ and the collusion resilience level $k$. When the codebook efficiency is fixed, higher resistance level requires longer fingerprinting codes. 
\end{itemize}

\subsection{Orthogonal Fingerprinting Evaluation}  \label{orthog_eval}
We evaluate the orthogonal fingerprinting methodology on MNIST dataset using a group of 30 users. As expected, orthogonal fingerprinting has good ability to distinguish individual users, while the collusion-resilience and scalability are not competitive with coded fingerprinting. In addition, orthogonal fingerprinting is essentially a special case of coded fingerprinting whose codebook is an identity matrix. For this reason, we evaluate the uniqueness, collusion resilience, and scalability of orthogonal fingerprinting while the comprehensive assessment is not shown here. 

\begin{figure}[ht!]
\centering
\includegraphics[width=0.4\textwidth]{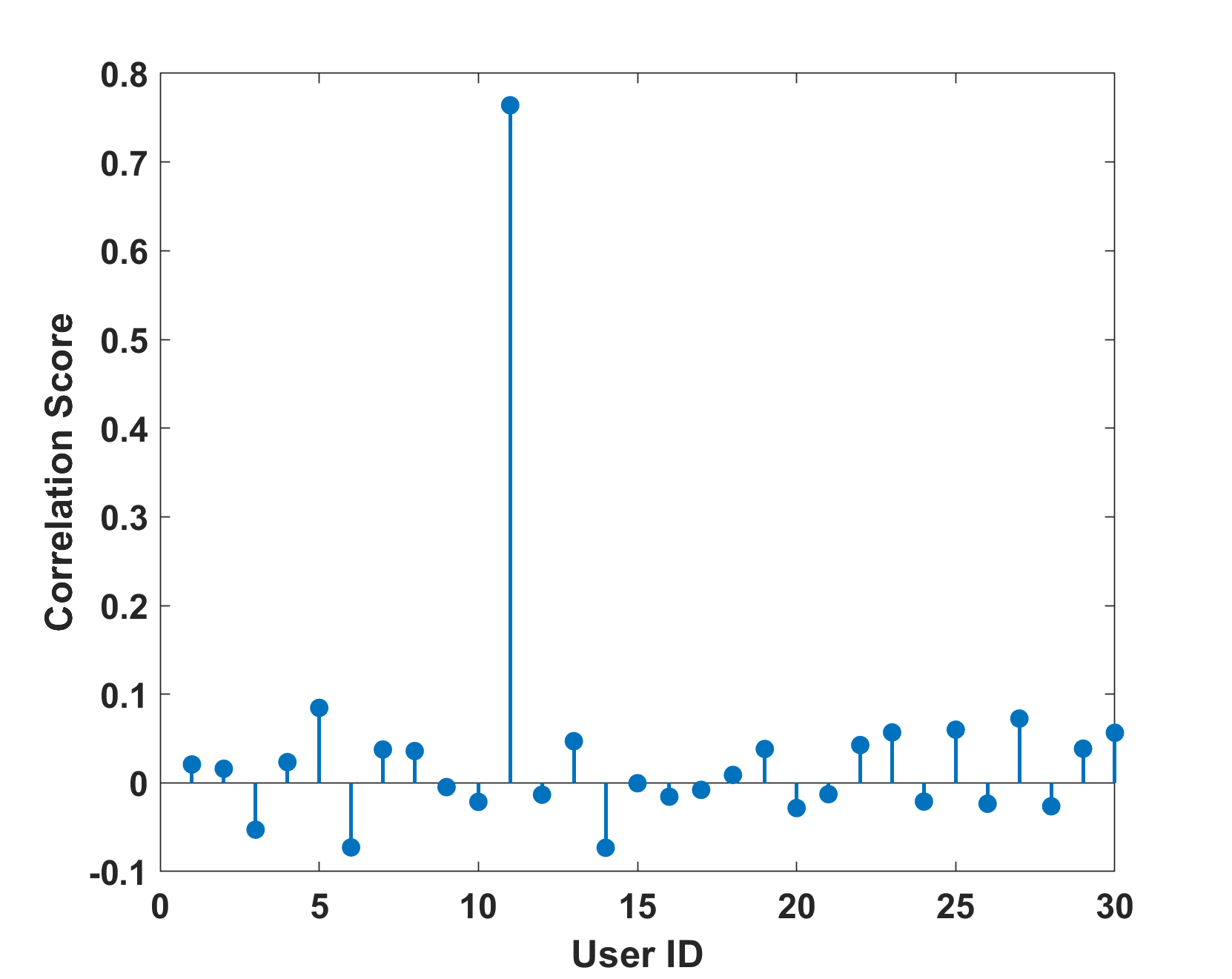} 
\caption{\label{fig:orthog_uniq_user11} Correlation scores of the fingerprint assigned to user 11 with the fingerprints of all 30 users when orthogonal fingerprinting is applied. The target user can be uniquely identified from the position of the ``spike'' in the correlation statistics. } 
\end{figure}

\vspace{0.5em}
\noindent \textbf{Uniqueness.} 
Each user can be uniquely identified by computing the correlation scores using Equation~\ref{eq:orthog_decode}. An example of user identification is shown in Figure~\ref{fig:orthog_uniq_user11} where user 11 is selected as the target. The correct user can be easily found from the position of the ``spike" in correlation scores due to the orthogonality of fingerprints.

\begin{figure}[ht!]
\centering
\includegraphics[width=0.4\textwidth]{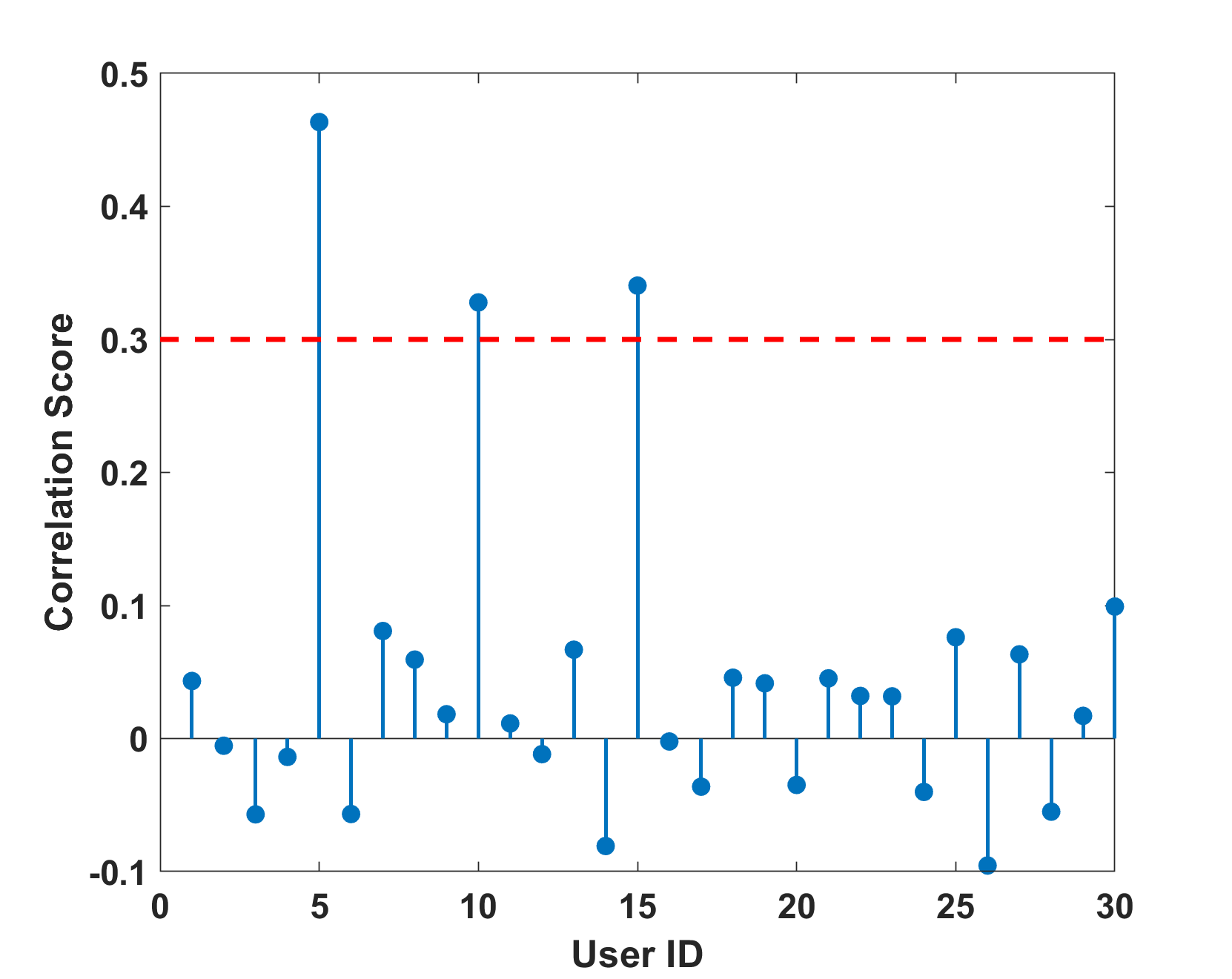} 
\caption{\label{fig:orthog_collusion_mnist_user5_10_15} Detection results of three colluders (user 5, user 10, and user 15) participating in the collusion attack. The red dashed line is the threshold (0.3) that can catch all colluders correctly. }
\end{figure}

\begin{figure}[ht!]
\centering
\includegraphics[width=0.4\textwidth]{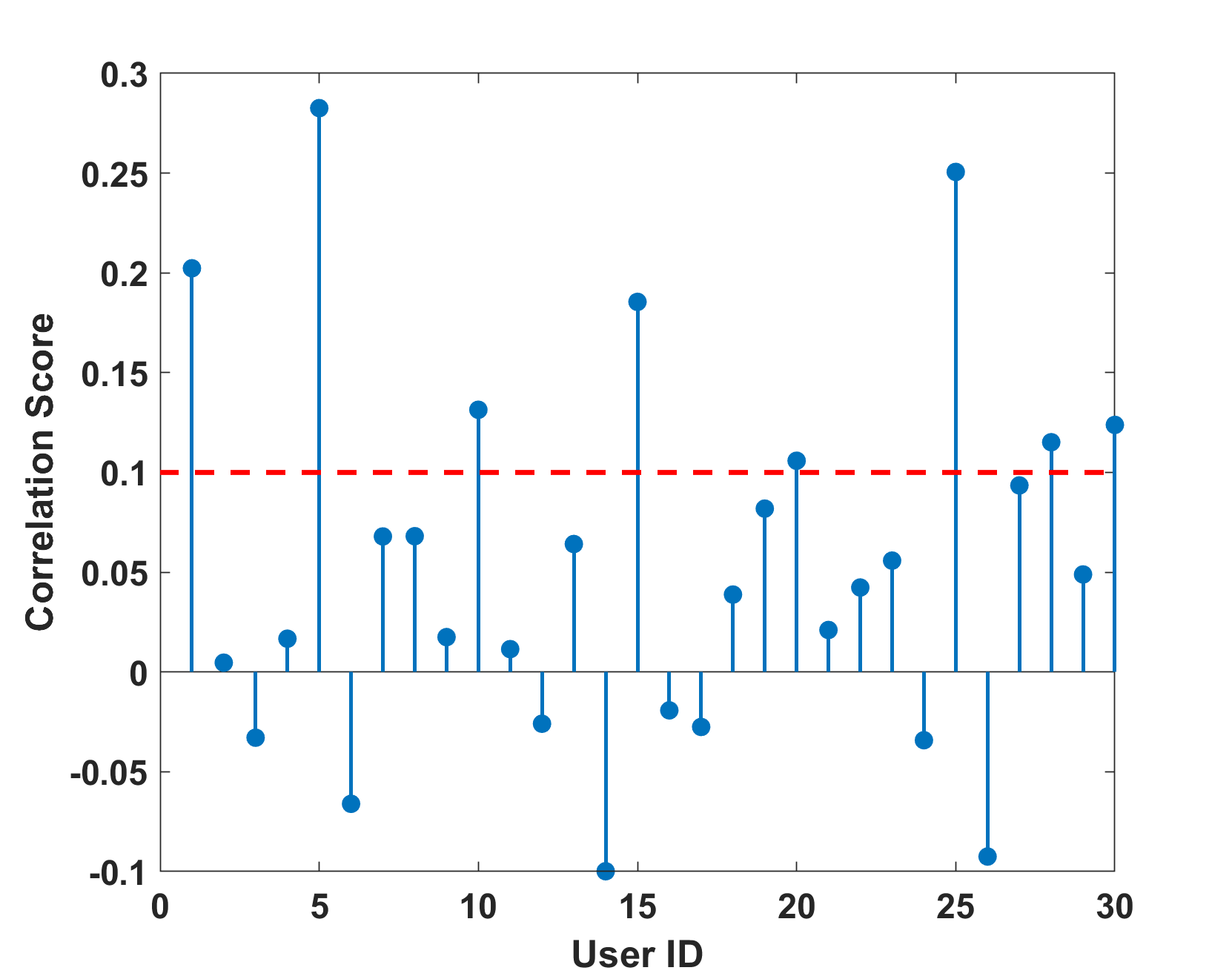} 
\caption{\label{fig:orthog_collusion_mnist_user1_5_10_15_20_25_28} Detection results of seven colluders (user 1, 5, 10, 15, 20, 25 and 28) participating in fingerprints collusion. User 30 is falsely accused if the red dashed line is used as the threshold. }
\end{figure}

\vspace{0.5em}
\noindent \textbf{Collusion resilience.} We evaluate the collusion resistance of orthogonal fingerprinting according to the colluder identification scheme discussed in Section~\ref{orthog_detect}. The detection results of three colluders are shown in Figure~\ref{fig:orthog_collusion_mnist_user5_10_15}, which suggests that the three participants in the collusion attack can be accurately identified by thresholding the correlation scores. However, when more users contribute to the collusion of fingerprints, the correlation scores of true colluders attenuate fast and the colluder set cannot be perfectly identified. Figure~\ref{fig:orthog_collusion_mnist_user1_5_10_15_20_25_28} shows the detection results of seven colluders where user 30 is falsely accused with the decision threshold denoted by the red dashed line. As can be seen from Figure~\ref{fig:orthog_collusion_mnist_user1_5_10_15_20_25_28}, there is no decision threshold that can ensure complete detection of all colluders and no false alarms of innocent users. Thus, orthogonal fingerprinting suffers from fingerprints attenuation and has a high chance of false positives as well as false negatives when collusion happens.

\vspace{0.5em}
\noindent \textbf{Scalability.} Orthogonal fingerprinting requires the code length to be $\mathcal{O}(n)$ bits for accommodating $n$ users, which could be much larger than the code bits $\mathcal{O}(\sqrt{n})$ needed in coded fingerprinting. Thereby, the scalability of orthogonal fingerprinting is inferior to that of code modulated fingerprinting.

\section{Conclusion}  \label{conclusion}
In this paper, we propose \sys{}, the first generic DL fingerprinting framework for IP protection and digital right management. Two fingerprinting methodologies, orthogonal fingerprinting and coded fingerprinting, are presented and compared. \sys{} works by embedding the fingerprints information in the probability density distribution of weights in different layers of a (deep) neural network. The performance of the proposed framework is evaluated on MNIST and CIFAR10 datasets using two network architectures. Our results demonstrate that \sys{} satisfies all criteria for an effective fingerprinting methodology, including fidelity, uniqueness, reliability, integrity, and robustness. \sys{} attains comparable accuracy to the baseline neural networks and resists potential attacks such as fingerprints collusion, parameter pruning, and model fine-tuning. The BIBD AND-ACC modulated fingerprinting of \sys{} has consistent colluders detection performance across benchmarks, suggesting that our framework is generic and applicable to various network architectures. 

% future work: optimize colluder detection scheme in coded fingerprinting, better codebook design....

\bibliographystyle{IEEEtran}
\bibliography{ref}
\end{document}